\documentclass[5p,10pt,a4paper]{elsarticle}
\usepackage{graphicx}
\usepackage{amssymb}
\usepackage{amsthm}
\usepackage{amsmath}
\usepackage[breaklinks]{hyperref}
\usepackage{xcolor}
\usepackage{lineno}
\usepackage{multirow}
\usepackage{bibentry}
\modulolinenumbers[1]
\usepackage{balance}
\usepackage{chemfig}
\usepackage[final]{pdfpages}

\journal{Materials Science and Engineering: A}

\begin{document}

\begin{frontmatter}

\title{{Nanoindentation 
of single crystalline Mo: Atomistic defect nucleation and  thermomechanical stability}}

\author[a,b]{F. J. Dom\'inguez-Guti\'errez\corref{author}}
\author[a]{S. Papanikolaou}
\author[a]{A. Esfandiarpour}
\author[a]{P. Sobkowicz}
\author[a,c]{M. Alava}

\cortext[author] {Corresponding author: javier.dominguez@ncbj.gov.pl}
\address[a]{NOMATEN Centre of Excellence, National Centre for Nuclear Research, ul. A. Sołtana 7, 05-400 Swierk/Otwock, Poland}
\address[b]{Institute for Advanced Computational Science, Stony Brook University, Stony Brook, NY 11749, USA}
\address[c]{Department of Applied Physics, Aalto University, P.O. Box 11000, 00076 Aalto, Espoo, Finland}

\begin{abstract}
The mechanical responses of single crystalline Body-Centered Cubic (BCC) metals, such as molybdenum (Mo), 
outperform other metals at high temperatures, so much so that they are considered as excellent candidates for applications under extreme conditions,
such as the divertor of fusion reactors. 
The excellent thermomechanical stability of molybdenum at high temperatures (400-1000$^{\rm o}$C) has also been  detected through nanoindentation,
pointing towards connections to emergent local dislocation mechanisms 
related to defect nucleation. 
In this work, we carry out a computational study of the effects of 
high temperature on the mechanical deformation properties of single
crystalline Mo under nanoindentation. 
Molecular dynamics (MD) simulations of spherical nanoindentation are 
performed at two indenter tip diameters and crystalline sample orientations
[100], [110], and [111], 
for the temperature range of 10-1000K. 
We investigate how the increase of temperature influences the 
nanoindentation process, modifying dislocation densities, mechanisms, 
atomic displacements and also, hardness, in agreement with reported
experimental measurements. 
Our results suggest that the characteristic formation and high-temperature
stability of [001] dislocation junctions in Mo during nanoindentation, 
in contrast to other BCC metals, may be the cause of the persistent
thermomechanical stability of Mo.

\end{abstract}


\end{frontmatter}

\section{Introduction}
\label{sec:intro}

The development of novel technology in aerospace, electronic, 
medical and energy industries requires the use of materials 
that can mechanically sustain extreme operating conditions, 
that may include, among others, high temperature and irradiation. 
In these environments, it is characteristic that BCC metals, 
such as tungsten and molybdenum, display excellent features. 
For example, molybdenum is a 
material with persistent high temperature hardness and strength, 
as well as high resistance to corrosion 
\cite{KeMin,2010M2009375,HOLLANG2001233}, and good 
thermodynamic properties at high pressure 
\cite{Litasov,doi:10.1063/1.324094,CIESZYKOWSKA2017124}.  
These properties have promoted molybdenum to be used 
for building a plasma wall component and fusion divertors over 
other materials
\cite{2010M2009375, LI2018367,doi:10.1098/rspa.1980.0043,BUDD1990129,IterDivertor}.
Nevertheless, compared to face-centered cubic (FCC) metals, 
the plastic behavior (strength, hardness) of BCC metals is quite 
complex and remains relatively unexplored, with only few nanoindentation studies addressing dislocation mechanisms in pure BCC single
crystals~\cite{Smith:2003aa,Durst:2006aa,Stelmashenko:1993aa,Syed-Asif:1997aa,Bahr:1998aa,Kramer:2001aa,Biener:2007aa}. 
In particular, it has remained poorly understood why some BCC 
metals display high thermomechanical stability at high temperatures, 
and persistence of  mechanical properties. 
Especially, Mo single crystals display persistent hardness at 
high temperatures~\cite{PlummerOxford}, a feature that is absent 
in other BCC single-crystalline metals~\cite{Biener:2007aa}. 
While the emergence of such thermomechanical stability has previously 
been attributed to grain-boundary effects~\cite{VOYIADJIS2010307}, 
in single crystals the effects shall be connected to fundamental 
dislocation mechanisms. 
Furthermore, the observation of corresponding thermomechanical 
stability in nanohardness, through nanoindentation, suggests the 
connection to defect nucleation~\cite{PlummerOxford}. 
Here, we investigate the mechanical nanoindentation response of Mo 
to a spherical indenter, at high temperatures, by using MD simulations, 
and considering [100], [110], and [111] orientations in a temperature 
range of 10-1000K, and a repulsive imaginary indenter. 
Our results suggest that junction formation is prevalent in Mo, 
especially compared to other BCC metals, such as Ta~\cite{Biener:2007aa} 
or W~\cite{Syed-Asif:1997aa,TERENTYEV2020105222,BEAKE201863,Pisarenko}.

Temperature effects on plasticity of Mo have been shown to span a range
of properties, from  yield stress to the resistance to 
fatigue and creep to restrict progressive deformation
\cite{VOYIADJIS2010307,BRAUN2019104999}. 
Ab initio simulations have been used for the study beyond the point
of maximum elastic deformation~\cite{PhysRevB.66.094110} and coupled
atomistic continuum methods have been extensively used to simulate 
nanoindentation of Mo, concluding that plasticity mechanisms are 
consistent with typical mechanisms observed in other BCC
metals~\cite{VOYIADJIS2010307,RPicu,10.1007/978-981-10-4109-9_34}. 
In the context of BCC metals, studies of nanoindentation in
Ta~\cite{PhysRevLett.109.075502} have shown that nanocontact 
plasticity in BCC is driven by the nucleation and propagation of 
twin and stacking fault bands, and Remington et al. showed that 
a cowboy-like "lasso" mechanism is responsible for the formation 
of prismatic loops \cite{Remington:2014aa}. 
In addition, the investigation of the temperature and loading-rate
dependence of the first pop-in load in nanoindentation of Ta has 
provided further light to collective dislocation
mechanisms~\cite{Sato:2019aa}. 
A complementary dimension has been added through uniaxial 
compression studies of micropillars that showed the importance of 
the relative sample orientation and also, the key role of screw
dislocations at room temperature in BCC metals (Ta, Mo and Fe), 
in contrast to FCC (Cu)~\cite{Kaufmann:2013aa,Kaufmann:2011aa}. 
Even though plasticity mechanisms in micropillar compression may be
different than nanoindentation and quite
complex~\cite{Papanikolaou:2017aa}, the key role of screw 
dislocations in driving plasticity in BCC metals cannot be 
understated.
In this context, Molybdenum and its alloys are tougher than W 
facilitating the manufacturing process for the fusion machine components
\cite{BUDD1990129}, where the mechanical properties such hardness 
and elastic modulus can be measured by nanoindentation 
\cite{cryst7100321,ZHAO2018365,VOYIADJIS2010307,PlummerOxford,MINNERT2020108727} 
enabling to 
analyze thermal activated mechanisms that change the 
mechanical properties of the material. 
However, nanoindentation experiments at elevated temperatures 
have several technical issues that need to be considered, e.g.
thermal drift that can be caused by
a temperature mismatch between the tip and the 
sample \cite{WHEELER2015354}.

{
A possible way to investigate features of nanomechanical properties and their causal relation to atomistic defect mechanisms at high temperatures, is through atomistic simulations. Molecular dynamics (MD) simulations have been extensively pursued in the past, showing to be a powerful tool towards emulating nanoindentation experiments, albeit at the nanoscale of few nanometers, while providing atomistic insights to the mechanical response of indented samples~\cite{10.1007/978-981-10-4109-9_34,PhysRevLett.109.075502,Sato:2019aa,10.1088/1361-651X/abf152,PLIMPTON19951,2007MJ200769}.
The major advantage of MD simulations at high temperatures is the ability to  investigate the thermomechanical stability of dislocation nucleation and propagation, as well as stacking faults and twin boundaries, and how they contribute to the increase or 
decrease of hardness~\cite{Lee}.
These atomistic simulations can also be applied to study the 
anisotropy of mechanical properties of materials at different 
temperatures which is a helpful tool to guide experiments 
where technical limits and costs are important 
\cite{Christopher_2001,Lee}.

In this work, we perform MD simulations to emulate nanoindentation 
of crystalline molybdenum by considering [100], [110], and 
[111] orientations in a temperature range of 10-800K, 
with a spherical indenter. In Section~\ref{sec:com_meth} we describe the details of the numerical simulations. In Section~\ref{sec:Results}, the atomistic insights of indentation processes in crystalline Mo sample are presented, where we track hardness as function of temperature, in conjunction to dislocation loop formation, local displacements' magnitude, and dislocation densities, showing the effects of sample temperatures on nanoindentation mechanisms. 
Agreement with experimental measurements is reported and discussed, in connection to dislocation junction formation. Finally, in section \ref{sec:Concl.}, we provide concluding remarks.
}

\section{Computational methods}
\label{sec:com_meth}

\begin{figure}[b!]
   \centering
   \includegraphics[width=0.40\textwidth]{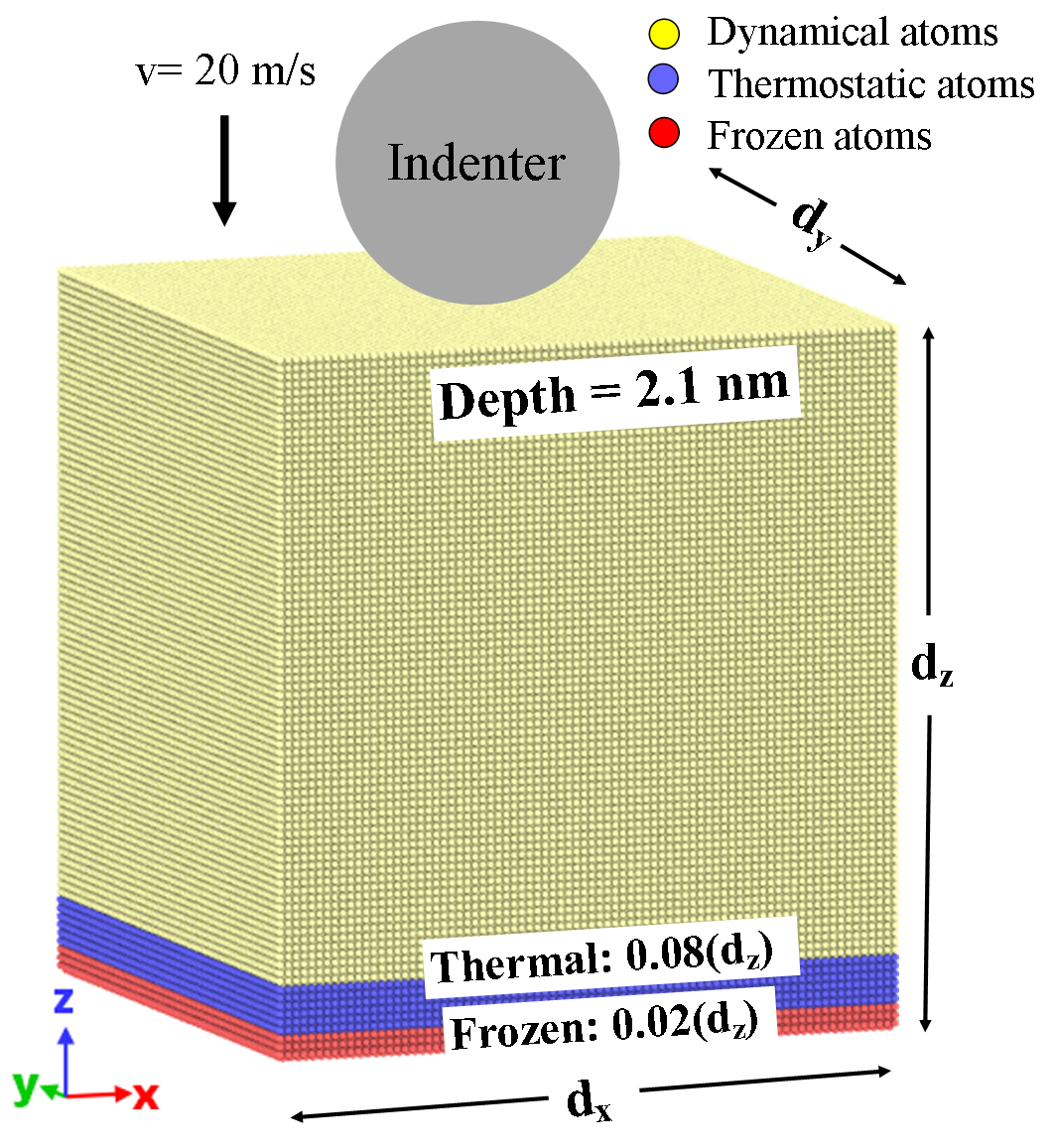}
   \caption{(Color on-line). Schematic of the standard configuration in our MD simulations of Mo nanoindentation. The prepared Mo sample is divided into three regions to consider boundary conditions and a non-atomic repulsive spherical indenter is used.}
   \label{fig:fig1}
\end{figure}

In order to perform MD simulations of nanoindentation, the 
prepared sample is divided into three sections on the $z$ direction 
for setting up boundary conditions, as shown in Fig. \ref{fig:fig1}. 
In addition, we consider 5 nm vacuum section 
on the top, above the material sample and also the two lowest bottom
layers are kept frozen ($\sim$ 0.02$\times$\textbf{d$_z$}) to 
assure stability of the Mo atoms when
nanoindentation is performed.
A thermostatic region above the already defined frozen one is 
considered to dissipate the generated heat during 
nanoindentation, with a thickness of $\sim$ 0.08$\times$\textbf{d$_z$}. 
The rest of the layers are defined as a region with dynamical atoms, 
in which the atoms interact as the indenter tip modifies the 
surface structure of the Mo sample. The dimensions of the simulation box, for the simulations performed in this work, are mentioned in Table~\ref{tab:MDdata}.

In the dynamical atoms region, the numerical modeling of a nanoindentation process starts by defining the simulation box as a pristine 
crystalline Mo sample based on a body-centered cubic cell. Then, we apply molecular dynamics through a NVE statistical thermodynamic ensemble where the velocity Verlet algorithm is implemented in the Large-scale Atomic/ Molecular Massively 
Parallel Simulator (LAMMPS) software
\cite{PLIMPTON19951,Frenkel:2001aa}
Periodic boundary conditions are set on the $x$ and $y$ axes to 
simulate an infinite surface, while the $z$ orientation contains 
a fixed bottom boundary and a free top boundary in all MD simulations.

\begin{table}[tbh]
    \centering
    \caption{Simulation boxes sizes and number of Mo atoms}
    \begin{tabular}{c c c}
       \hline
        \textbf{Orientation} & \textbf{Size} (\textbf{d$_x$}, 
        \textbf{d$_y$, \textbf{d$_z$}}) [nm] & \textbf{Atoms}  \\
        \hline
        $[001]$ &  25.18 $\hat{x}$  $\times$ 25.81 $\hat{y}$ 
        $\times$ 26.90 $\hat{z}$ & 1 128 320 \\
        $[110]$ &  24.80 $\hat{x}$ $\times$ 25.44 $\hat{y}$   
        $\times$ 28.71 $\hat{z}$ & 1 171 170 \\
        $[111]$ &  26.21 $\hat{x}$ $\times$ 25.81 $\hat{y}$ 
        $\times$ 30.10 $\hat{z}$ & 1 313 352 \\
        \hline
    \end{tabular}
    \label{tab:MDdata}
\end{table}

The indenter tip is considered as a non-atomic
repulsive imaginary (RI) 
rigid sphere with a force potential defined as: 
$F(t) = K \left(\vec r(t) - R \right)^2$ where $K = 236$ 
eV/\AA$^3$ (37.8 GPa) is the force constant, and $\vec r(t)$ is 
the position of the center of the tip as a function of time,
with radius $R$. Here, $\vec r(t) = x_0 \hat x + y_0 \hat y + (z_0 \pm vt)\hat z$ 
with $x_0$ and $y_0$ as the center of the surface sample on the 
xy plane, the $z_0 = 0.5$ nm is the initial gap between the surface 
and the intender tip moves with a speed $v$ = 20 m/s.
The loading and unloading processes are defined by considering 
the direction of the velocity as negative and positive, respectively.
Each process is performed for 125 ps with a time step of 
$\Delta t = 1$ fs. 
The maximum indentation depth is chosen to 2.1 nm 
to avoid the influence of boundary layers in the 
dynamical atoms region.

Our MD simulations are focused on standard nanoindentation simulations of body-centered cubic (BCC) molybdenum in the $[100]$, $[110]$, and $[111]$ crystal orientations. 
For this, we use the LAMMPS software and utilize the embedded-atom method 
(EAM) potential \cite{10.1088/1361-651X/abf152,AcklandGJ, Salonen_2003}. 
The Mo samples are initially energy optimized 
at 0 K by the conjugate gradient algorithm with energy 
tolerance of 10$^{-6}$ eV. The samples are then thermalized for 100 ps with a Langevin thermostat to temperatures of 10, 300, 600, 800 K and 1000K with the 
time constant of 100 fs. This is done until the system reaches a homogeneous 
sample temperature and pressure profile. A final step is performed by relaxing the prepared sample for 10 ps to dissipate artificial heat.
The elastic constants, C$_{ij}$, bulk modulus, 
and Poisson's ratio for different temperatures are computed by 
using the EAM potential for a small BCC Mo sample of 1 nm in the  
$z$ direction. The obtained values are presented in Table~\ref{tab:EC_Mo}.

\begin{table}[tbh]
\centering
\caption{
Elastic constants of molybdenum at various temperatures, obtained by MD simulations and compared with experimental values, in units of GPa. 
The lattice constant used in MD calculations is 3.1472 \AA{}, 
based on a body-centered cube unit cell.}
\begin{tabular}{c r r r r r r}
\hline
\textbf{C$_{ij}$} & \textbf{MD} & \textbf{Exp.} &
\textbf{MD} &
\textbf{MD}  & \textbf{MD} \\
\hline
 & \multicolumn{2}{c}{0 K} & 300 K & 600 K & 800 K\\
\hline
C$_{11}$  &  464.7   & 464 & 417.8  
& 354.1  & 321.6 \\
C$_{22}$  &  464.7   & 464 &  409.3    
& 340.5  & 307.6 \\
C$_{33}$  &  464.7   & 464 &  415.3
& 347.9  & 312.9 \\
C$_{12}$  &  161.5   & 159 &   163.2 
& 163.2  & 162.5 \\
C$_{13}$  &  161.5   & 159 &    163.3
& 164.7  & 165.1 \\
C$_{23}$  &  161.5   & 159 &     162.2
& 160.5  & 159.8 \\
C$_{44}$  &  108.9   & 109  &  108.4  
& 108.3  & 108.3 \\
C$_{55}$  &  108.9   & 109  &  107.0   
& 104.2  & 102.4 \\
C$_{66}$  &  108.9  &  109  &   106.1 
& 102.1  &  99.5 \\
\hline
Bulk Mod.  & 262.59 & 250   & 246.6 
& 224.40 & 212.96 \\
Poisson R. & 0.26 & 0.29   & 0.28 
& 0.32 & 0.34 \\
\hline
\end{tabular}
\label{tab:EC_Mo} 
\end{table}

\subsection{Calculation of mechanical properties}
\label{subsec:mech_prop}

The hardness of the indented sample is 
calculated by computing the $P-h$ curve with the Oliver and 
Pharr method \cite{oliver_pharr_1992}, following the fitting 
curve to the unloading process curve as: 
\begin{equation}
P = P_0 \left( h - h_f \right)^m
\label{Eq.OandP}
\end{equation}
with $P$ is the indentation load; 
$h$ is the indentation depth and $h_f$ is the residual depth 
after the whole indentation process; and $P_0$ and 
$m$ are fitting parameters. 
Thus, the nanoindentation hardness can be computed as: 
$H = P_{\rm max}/A_c$ where $P_{\rm max}$ is the maximum indentation 
load at the maximum indentation depth, $A_c = 
\pi \left(2R - h_c \right)h_c$ is the 
projected contact area with $R$ as the indenter tip radius 
and $h_c = h_{\rm max}- \epsilon P_{\rm max}/S$. 
Here $\epsilon = 0.75$ is a factor related to the spherical 
indenter shape, and unloading stiffness $S$ is calculated as
\begin{equation}
    S = \left( \frac{dP}{dh} \right)_{h = h_{\rm max}} 
    = mP_0 \left( h_{\rm max} - h_f \right)^{m-1}.
\end{equation}

The Young's module $E_{\rm Y}$ is computed as:
\begin{equation}
    \frac{1-\nu^2}{E_{\rm Y}} = \frac{1}{E_{\rm r}} - 
    \frac{1-\nu_{\rm i}^2}{E_{\rm i}},
\end{equation}
where $\nu$ and $\nu_{\rm i}$ are the Poisson's ratio of the 
Mo sample and indenter, respectively. 
$E_{\rm i}$ is the Young's modulus of the spherical indenter 
that is considered to be infinitely large, 
and the effective elastic modulus $E_{\rm r} = \sqrt{\pi/A_c}S/2\beta$ 
with $\beta = 1$ for a spherical indenter shape.
In this way, the nano-hardness of the indented samples 
can be calculated at different temperatures and 
indenter tip sizes.

The deformation of the indented Mo samples 
may be estimated through the use of the stress tensor $\sigma^n_{ij}$ of the 
$n$-th atom during the nanoindentation process. 
The von Mises stress of the $n$-th atom is
calculated as follows:
\begin{equation}
    \sigma^n_{\rm Mises} =   \sqrt{\frac{3}{2}s^n_{ij}s^n_{ij}}, 
    \quad 
    {\rm with} \quad
    s^n_{ij} =  \sigma^n_{ij}-\frac{1}{3}\sigma^n_{kk} \delta_{ij}. 
\end{equation}
The six components of $\sigma^n_{ij}$ are obtained from the output 
data of the MD simulations.

For the shear dependence of nanoindentation, atomic strains are computed through the 
distance difference, $\textbf{d}^{ \beta}$, between the the 
$m$-th nearest neighbors of the $n$-th atom of the 
pristine and indented samples. 
Followed by defining the Lagrangian strain matrix 
of the $n$-th atom as
\cite{2007MJ200769}:
\begin{eqnarray}
    \boldsymbol{\eta}_n = & 1/2 \left(\boldsymbol{J}_{n} \boldsymbol{J}_{n}^T-I\right), \\ 
      & {\rm with} \nonumber \\
    \boldsymbol{J}_{n} = & \left( 
    \sum_{m}\boldsymbol{d}_{m}^{0T}\boldsymbol{d}_{m}^{0} 
    \right)^{-1} 
    \left(\sum_{m}\boldsymbol{d}_{m}^{0T}
    \boldsymbol{d}_{m} \right).
\end{eqnarray}
Thus, the shear invariant of the $n$-th atom is computed as:
\begin{equation}
    \eta_n =   \sqrt{\frac{\zeta_{ij}\zeta_{ij}}{2}}, \quad 
    {\rm with} \quad
    \zeta_{ij} =  \eta_{ij}-\eta_{kk} \delta_{ij}. 
\end{equation}
This approach is implemented in OVITO \cite{ovito}. 

\section{Results}
\label{sec:Results}

\subsection{Temperature Dependence of nano-hardness for $[001]$, $[110]$, and $[111]$ orientations.}

\begin{figure}[b!]
   \centering
   \includegraphics[width=0.45\textwidth]{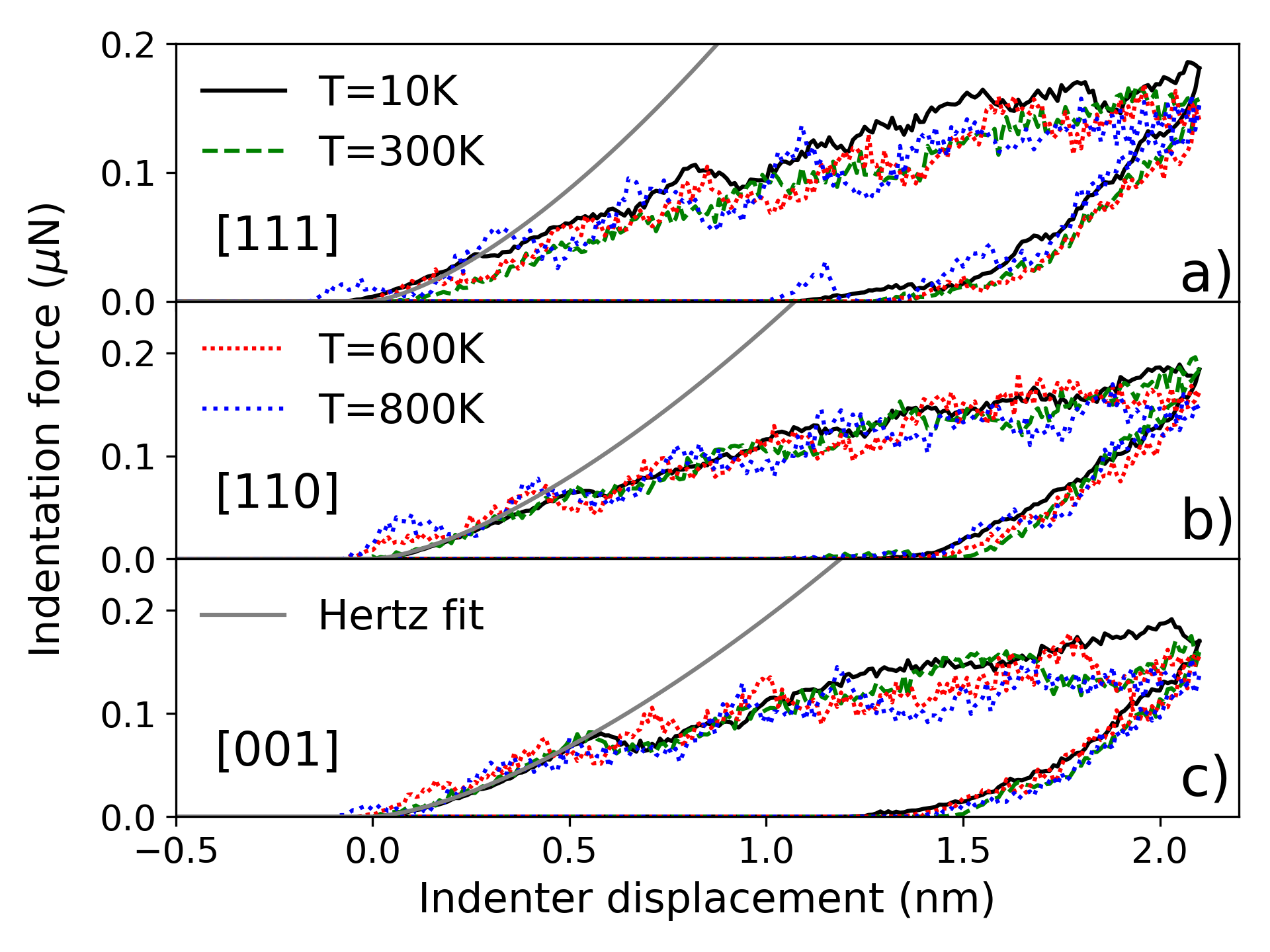}
   \caption{(Color on-line). $P-h$ curve of the nanoindentation of 
   crystalline Mo for the $[111]$ in a), $[110]$ in b), and $[001]$ 
   in c) crystal orientations at 10 to 800 K sample temperatures for indenter radius 3 nm. 
   A Hertz fitting curve, defined by Eq. \ref{Eq:Hertz}, is added to the results at 10 K.}
   \label{fig:Ph_3nm}
\end{figure}

The loading and unloading processes in nanoindentation of 
crystalline Mo oriented in $[001]$, $[110]$, and $[111]$ are 
recorded, and the P-h curves are presented in 
Fig. \ref{fig:Ph_3nm} for a tip radius of 3 nm (for larger tip radius (6 nm), please see Sec.~\ref{sub:indentertip}), at the different sample temperatures. At contact, all studied samples follow classical Hertz 
theory where the sphere-flat surface contact solution is  expressed as
\begin{equation}
    P^{hkl} = \frac{4}{3} \frac{E^{hkl}_Y}{1-\nu^2}R^{1/2}h^{3/2},
    \label{Eq:Hertz}
\end{equation}
where $hkl$ are indexes related to the crystal orientation, 
$\nu$ is the Poisson's ratio, $R$ is the indenter radius, 
$h$ is the indenter displacement, and $E^{hkm}_Y$ is 
the elastic modulus calculated as
\begin{eqnarray}
    \frac{1}{E^{hkl}_Y} = & S_{11} -2\left[\left(S_{11}-S_{12}\right)
    -\frac{S_{44}}{2} \right] \nonumber \\
    & \times \left(\alpha^2 \beta^2 + 
    \alpha^2 \gamma^2 + \beta^2 \gamma^2\right),
\end{eqnarray}
with $S_{ij}$ as the components of the compliance tensor and 
$\alpha$, $\beta$, and $\gamma$ are the direction
cosines of the $[hkl]$ crystal orientation. 
Obtaining a elastic moduli of $E^{001}_Y = 381.39$ GPa, 
$E^{110}_Y$ = 287.02 GPa, and $E^{111}_Y$ = 305.94 GPa at the lowest 
temperature.
It is observed that the deviation of MD simulations  
from the Hertzian elasticity 
solution at 10K takes place at different deformations, greater 
than  0.57 $\pm$ 0.05 nm for $[001]$, 
0.5 $\pm$ 0.05 nm for $[110]$, and 0.26 $\pm$ 0.05 for the
$[111]$ crystal orientation, consistent with experimental findings of orientation dependence in crystal plasticity of BCC metals~\cite{BEAKE201863,PhysRevLett.109.075502,Kaufmann:2013aa}.

\begin{table*}[!t]
    \centering
    \caption{Results obtained from MD simulation at the maximum
    indentation depth. Hardness, Young's modulus and yield stresses 
    are presented in units of GPa. 
    The mean pressure $\langle p_{\rm m} 
    \rangle$ is computed as the instance force divided by the contact
    area, A$_c$.}
    \begin{tabular}{c || r r r r | r r r r | r r r r}
       \hline
       Orient. & \multicolumn{4}{c}{ \textbf{[001]}} & 
     \multicolumn{4}{c}{\textbf{[110]}} & \multicolumn{4}{c}{ \textbf{[111]}}  \\
       \hline
           Temp. [K] & 10 & 300 & 600 & 800  
           & 10 & 300 & 600 & 800 
           & 10 & 300 & 600 & 800 \\
                   \hline
       P$_{\rm max}$ [$\mu$N] & 
       0.170 & 0.160 & 0.155 & 0.136 &
       0.184 & 0.182 & 0.161 & 0.148 & 
       0.181 & 0.160 & 0.158 & 0.152 \\
       A$_{\rm C}$ [nm$^2$] & 
       23.85 & 24.06 & 23.83 & 23.97     &
       23.41 & 24.15 & 23.81 & 23.96    & 
       23.72 & 24.09 & 24.00 & 23.33 \\
       S[N/m] & 
       432.5  & 461.76 & 387.87 & 424.42 & 
       369.3  & 504.54 & 403.27 & 301.29 & 
       416.9  & 456.27 & 399.8 & 317.81 \\
        H & 
        6.92  & 6.59 & 6.28 & 6.38 &
        7.23  & 6.84 & 6.60 & 4.57 & 
        7.12  & 6.40 & 5.89 & 6.43 \\
        $E_Y$  & 
        73.17  & 76.87 & 63.20 & 67.94 & 
        63.06  & 84.83 & 65.75 & 48.25 & 
        70.74  & 76.81 & 64.91 & 51.57 \\
        \hline
        $\sigma_{\rm Hydro.}$ & 
        -5.63  & -6.23 & -9.08 & -10.79 & 
        -6.67  & -7.20 & -10.09 & -14.31 & 
        -5.16  & -5.27 & -8.38 & -12.39  \\
        $\sigma_{\rm Mises}$ & 
        4.81  & 3.76 & 3.71 & 6.79 & 
        4.20  & 3.06 & 4.01 & 6.71 & 
        2.86  & 3.12 & 3.72 & 7.00 \\
        $\sigma_{\rm Tresca}$ & 
        2.51  & 2.06  & 2.06 & 3.42 & 
        2.29  & 1.65 & 2.25 & 3.61 &
        1.49  & 1.79 & 1.92 & 3.57 \\
        $\langle p_{\rm m} \rangle$ & 
        7.13  & 6.63 & 6.50 & 5.71 &
        7.86  & 7.53 & 6.76 & 6.18 &
        7.63  & 6.64 & 6.58 & 6.51 \\
        $\frac{\sigma_{\rm Tresca}}{\langle p_{\rm m} \rangle}$ & 
        0.35  & 0.31 & 0.32  & 0.59 & 
        0.29  & 0.22 & 0.33 & 0.51 &
        0.19  & 0.27 & 0.29 & 0.55 \\
        \hline
    \end{tabular}
    \label{tab:MDdata3nm}
\end{table*}

We notice that the residual depth after indentation 
is h$_f = 1.4$ $\pm$ 0.1 nm for [100] and [110] orientations, 
while a value of h$_f = 1.0$ is found for the [111] orientation 
at 10 K. 
When the sample is thermalized to 600K, the elastic section of 
the P-h curve starts to oscillate due to the deformation of the 
surface shifting the residual depth after unloading process. 
This effect is more relevant at a temperature of 800K, 
where the Mo atoms interact with the spherical indenter before 
touching the surface. 
We also notice that the indentation force at 
 maximum depth, decreases as a function 
of the temperature, regardless of crystal orientations, in good 
agreement with reported results for Ta~\cite{PhysRevLett.109.075502}. 
The plastic region of the curves show more oscillations at 
higher temperatures due to the thermal motion of dynamical atoms region
in the sample.

The resulting P-h curve for the unloading process can be fitted to the 
Hertz elastic solution according to Eq. \ref{Eq.OandP} 
to compute mechanical properties of the Mo sample like Hardness 
and Young's modulus, that should in principle compare to experimental data at comparable depths. 
In Tab. \ref{tab:MDdata3nm}, we report the following values that are 
computed at the maximum indentation depth: maximum force, 
P$_{\rm max}$ in $\mu$N, the projected contact area, A$_{\rm C}$ 
in nm$^2$, the stiffness, S in units of N/m, and Hardness and 
Young modulus in units of GPa.
As well as the hydrostatic stress $\sigma_{\rm Hydro.} = \left( 
\sigma_{xx} + \sigma_{yy} + \sigma{zz} \right)/3$; the total value of 
von Mises and Tresca stresses; the mean pressure $\langle p_m 
\rangle$ is computed as the instance force divided by the contact
    area, A$_c$, namely F$_{\rm max}$/A$_{\rm C}$.
We also verify the relation factor between the mean pressure and 
the maximum shear stress underneath the indenter, where the classical 
Hertz theory suggests a value of 0.465 (See Table~\ref{tab:MDdata3nm}). 
Our MD simulations corroborates that this factor depends on the atomic 
scale and sample temperature, as reported in the literature
\cite{GOEL2015249}.

\begin{figure}[!b]
   \centering
   \includegraphics[width=0.48\textwidth]{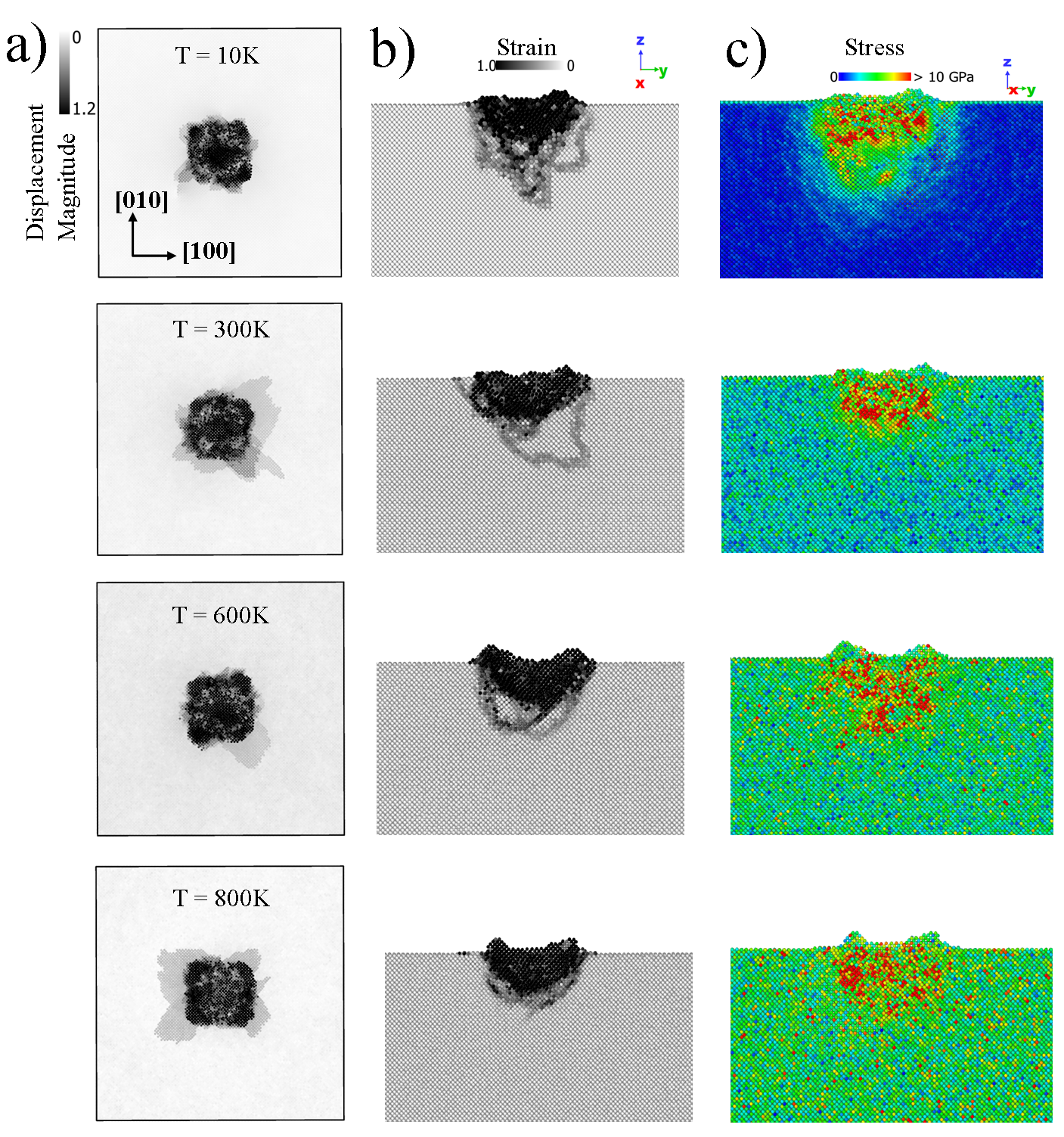}
   \caption{(Color on-line). Atomic analysis of the indented Mo 
   sample at different temperatures at a depth of 2.1 nm for 
   the $[100]$ crystal orientation. 
   Distribution of atomic displacement magnitude (in a range of 0-1 nm) 
   on the top view is shown in a). 
   von Mises strain and stress beneath the indenter are presented in b) 
   and c), respectively.
   }
   \label{fig:maxload3nm}
\end{figure}

In order to provide more insights about the effect of  
temperature in the mechanical response of Mo samples, we 
analyze atomic strain information. This analysis provides information 
about the topography of  surface deformation and pileup that 
can be compared to experimental results. 
In Fig. \ref{fig:maxload3nm}a), we present the distribution of 
atomic displacement magnitude of  
damage of indented Mo samples in a top view for the 
$[001]$ orientation with different sample temperatures, as an example. 
Noticing that pileups are found along the in-plane
slip directions \{101\} and \{10-1\}, this region is gradually enlarged 
by increasing the temperature due to atomic thermal 
motion, commonly identified in BCC investigations~\cite{PhysRevLett.109.075502}.

In Fig. \ref{fig:maxload3nm}b) and c), we report the atomic shear 
strain and von Mises stress for different temperatures and the $[001]$ 
orientation.
The sample is slid to the half in the $x$ axis to visualize 
the atomic distribution of the shear strain and von Mises stress 
by coloring Mo atoms according to their values.
The increase of temperature suppresses propagation 
of dislocation lines, and suppresses the size of the plastic zone.
It is also noticed that the strain and stress underneath the indenter 
increases as long as Mo atoms start moving faster due to 
thermal motion, which is also noted in the broader stress distribution.
The analysis applied to the $[110]$ and $[111]$ crystal orientations
are presented in the supplementary material (\ref{sec:appendix}). 
It is worth pointing out that the formation of a prismatic loop is observed 
on the [111] orientation at high temperatures, possibly suggesting that mechanisms shown at low temperatures~\cite{Remington:2014aa} are relevant also at high temperatures.

\begin{figure}[!b]
   \centering
   \includegraphics[width=0.48\textwidth]{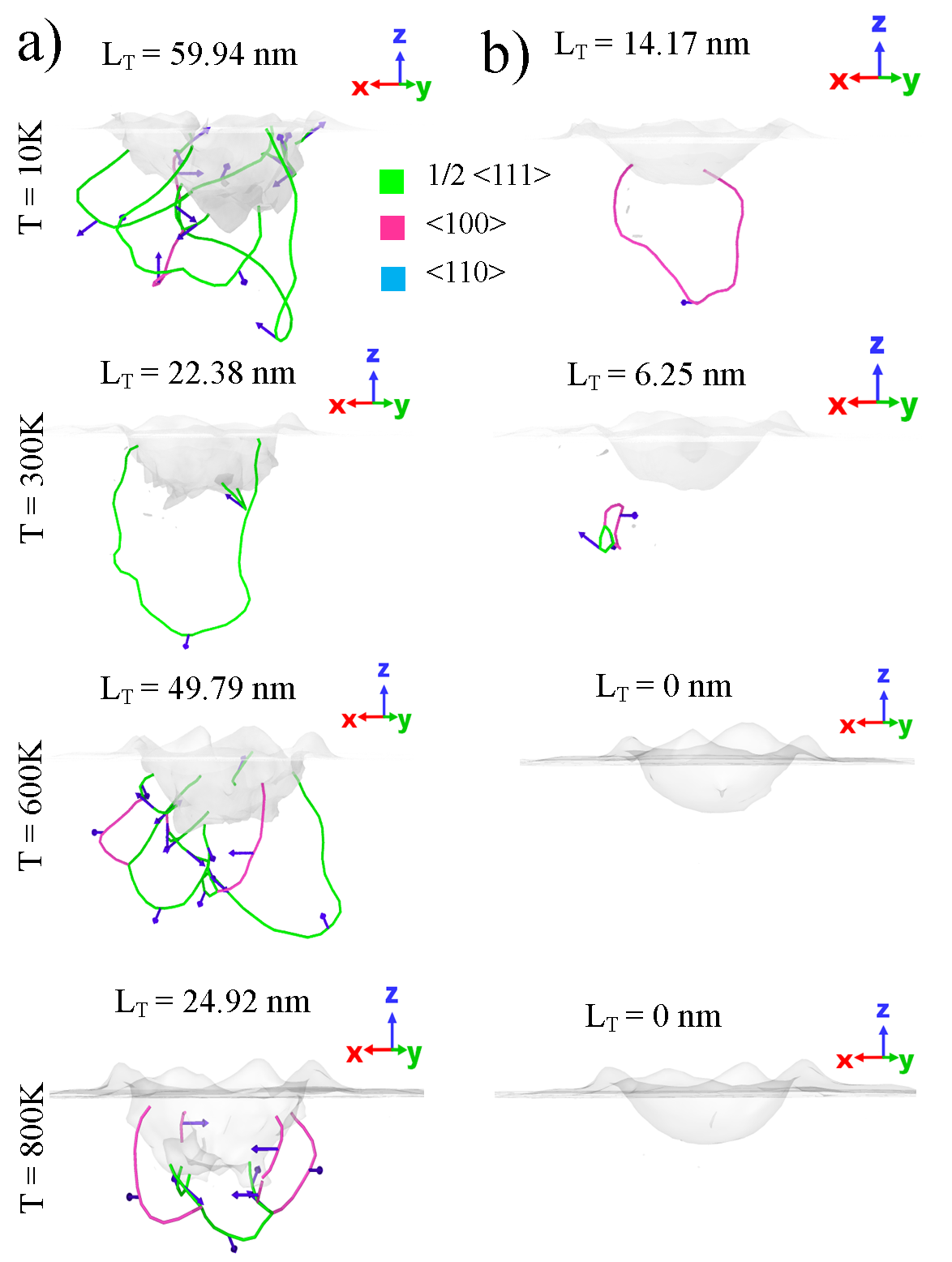}
   \caption{(Color on-line). Dislocation loops as a function of the indentation depth on the [001] orientation for different temperatures 
   at the maximum indentation depth in a) and after unloading 
   process in b). 
   We also include information of the total length of the dislocation 
   (L$_{\rm T}$).
   Different types of dislocation loops are colored in green 
   for $1/2 \langle 111 \rangle$, magenta for $\langle 100 \rangle$, 
   and blue for $\langle 110 \rangle$. 
   Burgers vector are displayed by blue arrows.}
   \label{fig:figloops3nm}
\end{figure}

It is interesting to analyze the atomic structure at the 
maximum indentation depth, where the computation and visualization 
of the dislocation structure are done by using 
the Dislocation Extraction Analysis (DXA) tool~\cite{ovito}. 
The output data provides the total length of dislocation 
lines and loops and their corresponding Burgers vectors. 
In Fig. \ref{fig:figloops3nm}a), we present the formation of 
dislocation loops at the maximum nanoindentation depth in the 
$[001]$ crystal orientation at a depth of 2.1nm. In Fig. \ref{fig:figloops3nm}b), the visualization of the deformed contact surface after unloading is shown, displaying the formation of pile-ups. 
We also include the information for the formation of 
shape and  Burgers vector (depicted as 
blue arrows) of the 
$1/2 \langle 111 \rangle$ (colored in green), 
$\langle 100 \rangle$ (colored in magenta), and 
$\langle 110 \rangle$ (colored in blue) dislocations. 
The total 
length of the dislocations, L$_{\rm T}$, is also noted, pointing to a dependence 
on the sample temperature after nanoindentation. 
In Fig. \ref{fig:figloops3nm}b) we display the remained dislocations 
after unloading process where the effect of temperature is 
observed through the quick adsorption of  dislocations on the sample surface. 
The effect of crystal orientation is reported in the 
supplementary material (\ref{sec:appendix}), 
where it is also shown that  dislocations disappear after 
nanoindentation loads are removed at high temperatures. 

\subsection{Effect of the indenter tip radius}
\label{sub:indentertip}

\begin{table*}[!h]
    \centering
    \caption{Results obtained from MD simulation at the maximum
    indentation depth with an indentation size of 6 nm.}
    \begin{tabular}{c || r r r r | r r r r | r r r r}
       \hline
       Orient. & \multicolumn{4}{c}{\textbf{[001]}} & 
     \multicolumn{4}{c}{\textbf{[110]}} & \multicolumn{4}{c}{ \textbf{[111]}}  \\
       \hline
           Temp. [K] & 10 & 300 & 600 & 800 
           & 10 & 300 & 600 & 800 
           & 10 & 300 & 600 & 800 \\
                   \hline
       P$_{\rm max}$ [$\mu$N] & 
       0.440 & 0.430 & 0.403 & 0.361 &
       0.430 & 0.382 & 0.423 & 0.373 & 
       0.399 & 0.359 & 0.331 & 0.361 \\
       A$_{\rm C}$ [nm$^2$] & 
       54.86 & 55.65 & 58.01 & 57.28    &
       46.29 & 51.84 & 52.37 &  55.14   & 
       54.93 & 55.13 & 55.93 & 55.55 \\
       S[N/m] & 
       781.37  & 849.71 & 1018.02 & 952.41 & 
       595.94  & 825.21 & 891.47 & 939.05 & 
       709.76  & 789.44 &  765.44 & 834.71 \\
        H & 
        7.70   & 7.65 &  6.72 & 6.97 &
        12.21  & 11.77 & 11.27 & 8.97 & 
        8.15   & 7.55 & 7.35 & 7.61 \\
        $E_Y$  & 
        87.17  & 93.03 & 106.32 & 109.06 & 
        72.38  & 96.71 & 97.99 & 99.12 & 
        80.27  & 86.83 &  89.58 & 87.78 \\
        \hline
        $\sigma_{\rm Hydro.}$ & 
        -15.66  & -17.12 & -19.63 & -19.90 & 
        -11.38  & -14.82 & -17.07 & -21.49 & 
        -11.62  & -9.95 & -14.33 & -12.38  \\
        $\sigma_{\rm Mises}$ & 
        7.58  & 4.54 & 1.26 & 4.05 & 
        8.16  & 5.69 & 2.49 & 4.53 & 
        8.33  & 5.71 & 1.71 & 7.00 \\
        $\sigma_{\rm Tresca}$ & 
        3.92  & 2.45 & 0.70 & 2.14 & 
        4.27  & 3.16 & 1.42 & 2.47 &
        4.36  & 2.92 & 0.99 & 3.58 \\
        $\langle p_{\rm m} \rangle$ & 
        8.02  & 7.72 & 6.94 & 6.30 &
        9.29  & 8.40 & 8.08 & 6.76 &
        7.47  & 6.45 & 6.99 & 6.50 \\
        $\frac{\sigma_{\rm Tresca}}{\langle p_{\rm m} \rangle}$ & 
        0.49  & 0.32 & 0.1  &  0.34 & 
        0.46  & 0.38 & 0.18  & 0.36 &
        0.58  & 0.45 & 0.17  & 0.55 \\
        \hline
    \end{tabular}
    \label{tab:MD_data_6nm}
\end{table*}

\begin{figure}[!b]
   \centering
   \includegraphics[width=0.5\textwidth]{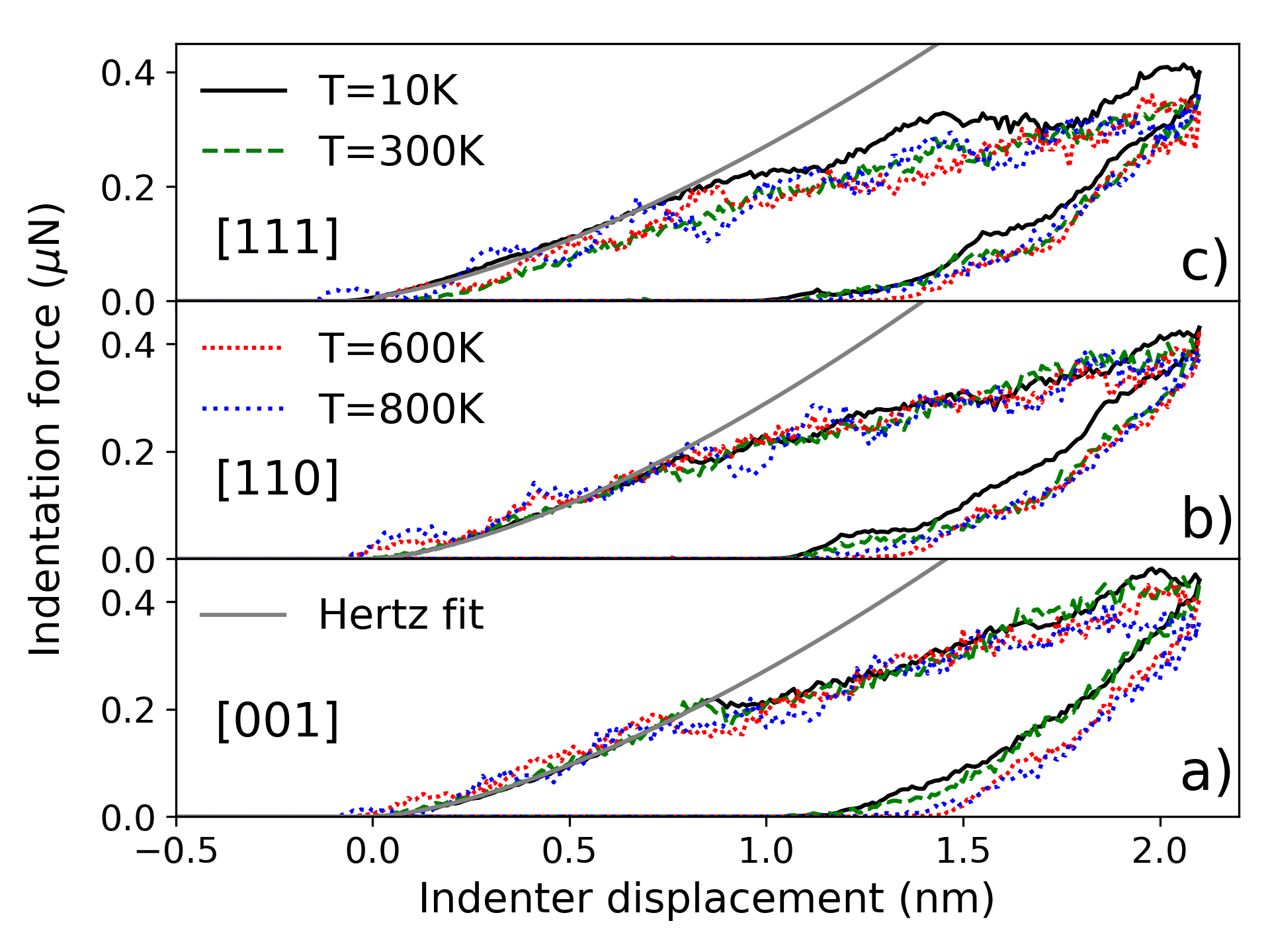}
   \caption{(Color on-line). 
   $P-h$ curve of the nanoindentation of 
   crystalline Mo at a sample temperatures of 10K to 800K for the $[001]$, 
   $[110]$, and $[111]$ orientations for a indented tip radius of 
   6 nm.
   A Hertz fitting curve is added to graphs for both cases.
   Temperature effects on mechanical nanoindentation response 
   are observed during the unloading process.
   }
   \label{fig:Ph_6nm}
\end{figure}

The spherical-tip nanoindentation studied in this work may resemble Berkovich nanoindentation at ultra-short depths ($<10$nm), however, in this regime, the mechanical response may display strong size effects~\cite{Bolin:2019aa}, naturally influencing relevant plasticity mechanisms. The effects of the indenter tip size on the observed 
nanoindentation plasticity mechanisms
of Mo samples, are tracked by considering a larger tip radius of 6 nm, 
followed by performing MD simulations under the same 
numerical environment as smaller tip ones.
In Fig.~\ref{fig:Ph_6nm}, we present the P-h curve of the 
nanoindetation process at different sample temperatures and 
crystal orientations. 
The corresponding classical Hertz solution (Eq. \ref{Eq:Hertz}) 
is superposed to show the effects of the 
temperature during the interaction of the surface and the 
sphere.
For this case, the MD simulations deviated from the Hertz elastic 
solution at 10K at 0.86 $\pm$ 0.05 nm for $[001]$, 
0.78 $\pm$ 0.05 nm for $[110]$, and 0.37 $\pm$ 0.05 for the
$[111]$ crystal orientation, also displaying very strong orientation dependence, as expected for BCC plasticity mechanisms~\cite{Kaufmann:2013aa}.
Note that the first pop-in load is independent of the 
indenter size, leading to the [001] crystal orientation 
to reach the maximum indentation force and the [111] one to the 
lowest value regardless the sample temperature. 
Nevertheless, the pop-in load decreases from the [110] to [111] and 
[001] at temperatures above 600 K. 
We also note that the residual depth is  1.1 $\pm$ 0.05 nm 
for the three crystal orientations at 10K. 
This value increases as function of the sample temperature.
In Tab. \ref{tab:MD_data_6nm}, we report the obtained hardness $H$, 
Young modulus $E_Y$, stiffness $S$, and yield stresses $\sigma$ as  
function of temperature for reference and further discussion.

\begin{figure}[!b]
   \centering
   \includegraphics[width=0.48\textwidth]{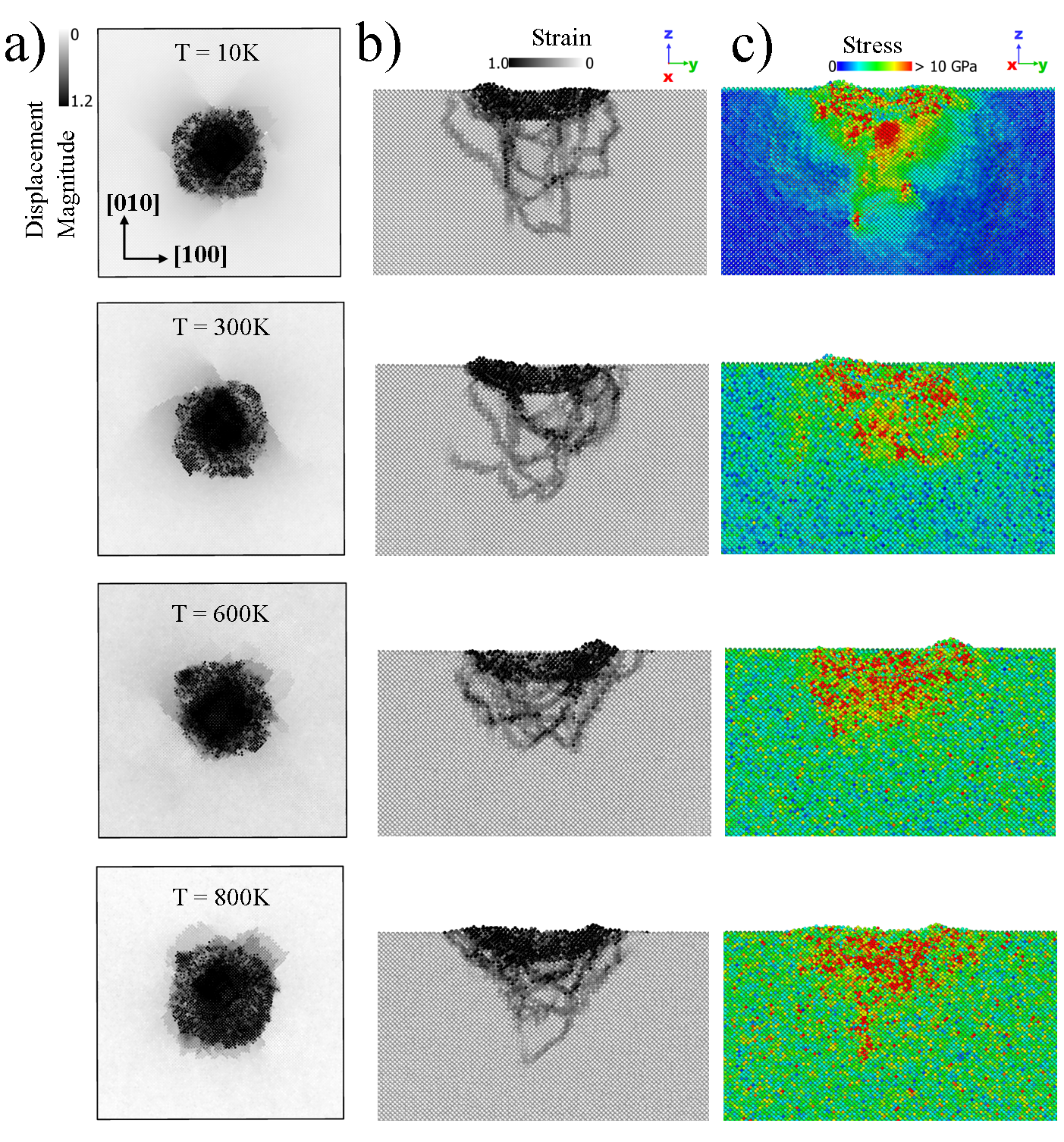}
   \caption{(Color on-line) 
   Analysis of the indented Mo sample for 
   the [001] crystal orientation as function of the temperature by 
   using a tip radius of 6 nm.
   Distribution of the atomic displacement magnitude is reported 
   in a), while von Mises strain and stress are shown in b) and c), 
   respectively.}
   \label{fig:maxload6nm}
\end{figure}

In Fig. \ref{fig:maxload6nm}, we present the resultant atomic 
configuration of Mo samples after loading, as function 
of temperature for the [100] orientation, as an example. 
The analysis for the indented of Mo samples at [110] and [111] 
orientations are presented in the supplementary material of 
this work (\ref{sec:appendix}).
In Fig. \ref{fig:maxload6nm}a), we report the distribution 
of the atomic displacement magnitude in a scale of 0 to 1.2 nm. 
Here, the pileup pattern at 10 K shows expectedly 
that the maximum shear stress is observed for the \{101\} and 
\{10-1\} slip 
systems, as also reported for single crystalline Mo samples
experimentally studied by Plummet et al. \cite{PlummerOxford}. 
This stress pattern is maintained at higher temperatures.
Fig. \ref{fig:maxload6nm} shows the atomic strain and von Mises
stress \ref{fig:maxload6nm}b) and c), respectively. 
While the thermally induced plastic zone reduction, observed for the smaller tip, is not present for the larger tip, there are close similarities to the atomic strain profiles and maxima as well as the temperature dependence of the von Mises stress.

\begin{figure}[!b]
   \centering
   \includegraphics[width=0.48\textwidth]{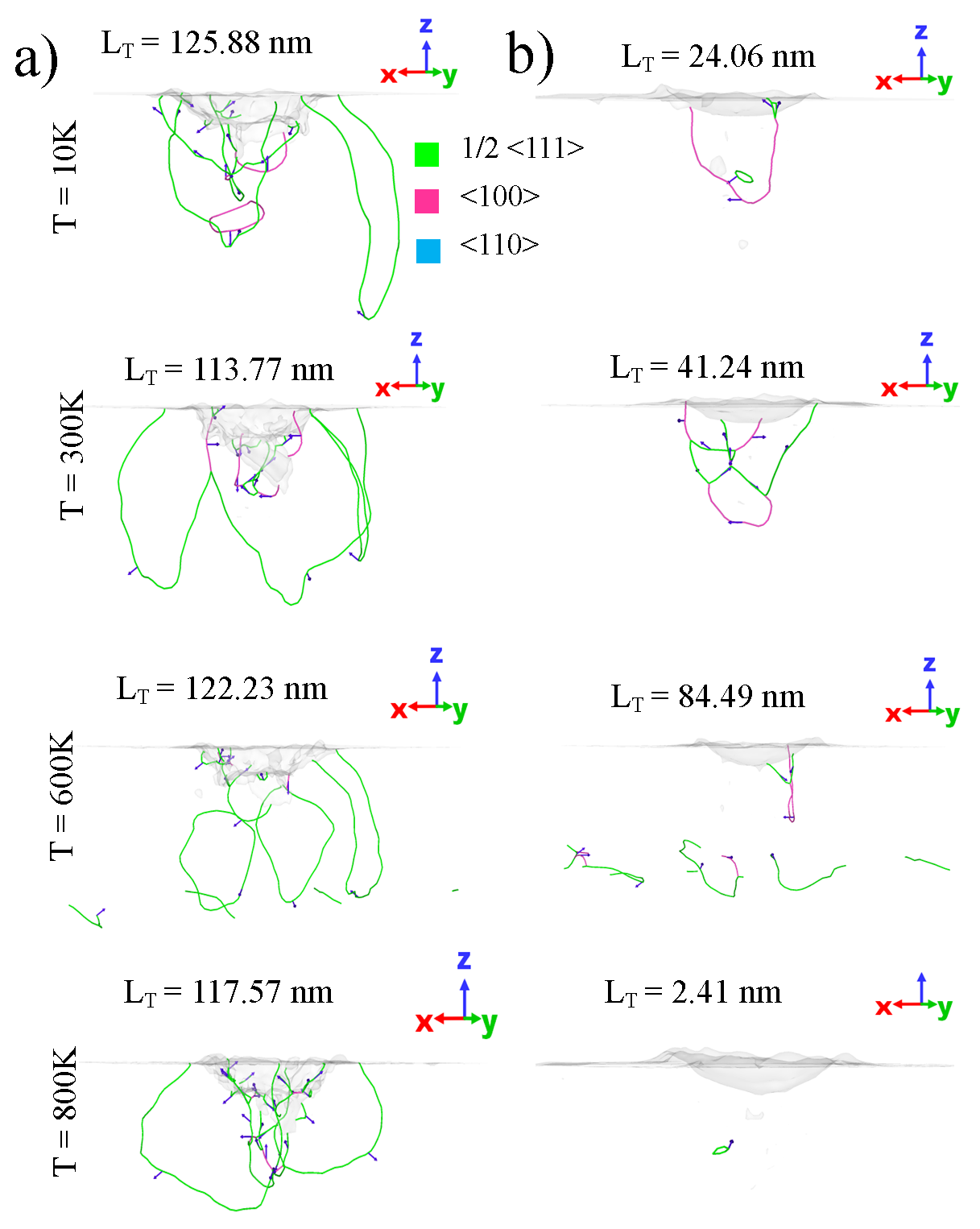}
   \caption{(Color on-line). Dislocations after maximum loading in a) 
   and after unloading process in b) for Mo samples on the 
   [100] orientation at different temperatures. 
   The total length of the dislocation (L$_{\rm T}$) is also presented.
   A radius of 6nm is considered for the nanoindentation process. 
   Color pattern follows the one used in Fig. \ref{fig:figloops3nm}.}
   \label{fig:Loops6nmTemp}
\end{figure}

In Fig. \ref{fig:Loops6nmTemp}, the plastic zone is characterized in two limits: a) at the maximum 
indentation depth and also, b) after nanoindentation, and it is  
characterized through the identification of the types of emerging dislocations at 
different temperatures. 
During loading, it is observed that the total length 
of the dislocations is larger than those reported for a smaller 
indenter regardless of the sample temperature. Evidence of the `lasso` mechanism~\cite{Remington:2014aa} exists at all temperatures, suggesting that the main dislocation nucleation mechanism remains analogous to other BCC metals.
After unloading, there is evidence of remaining dislocations that provide clues for the hardening mechanisms at high temperatures. 
Only at 800 K the dislocation nucleation appears to be overwhelmed by thermal atomic motions that drive dislocations towards surface deposition events.
The formation of prismatic loops is favorable for large radius 
indenters at all temperatures (up to 800K), justifying prior studies for other BCC metals.
\subsection{MD insights for experiments in Mo}
\label{subsec:disc}

\begin{figure}[!b]
   \centering
   \includegraphics[width=0.48\textwidth]{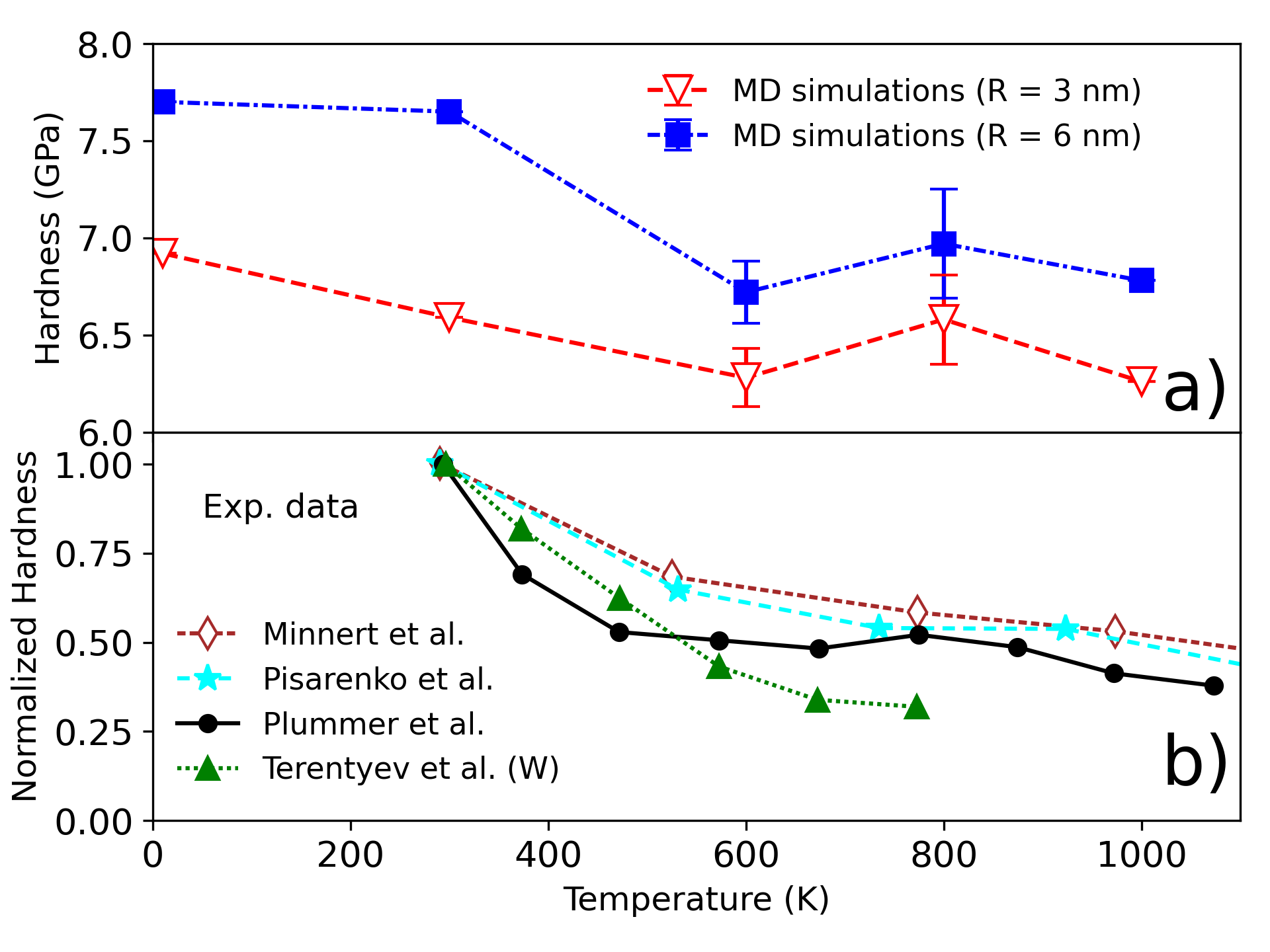}
   \caption{(Color on-line). Hardness as a function of temperature 
   for the [100] orientation at a depth of 2.1 nm 
   in a) and comparison to experimental  
   measurements in b) reported by Plummer et al. \cite{PlummerOxford}, 
   Pisarenko et al. \cite{Pisarenko}, Minnert et al. \cite{MINNERT2020108727}. 
   For comparison, we show experimental data 
   for W by Terentyev et al. \cite{TERENTYEV2020105222} showing 
   a monotonic decrease of the material hardness as function of 
    temperature.
   }
   \label{fig:ExpCompHardness}
\end{figure}

\begin{figure}[!b]
   \centering
   \includegraphics[width=0.48\textwidth]{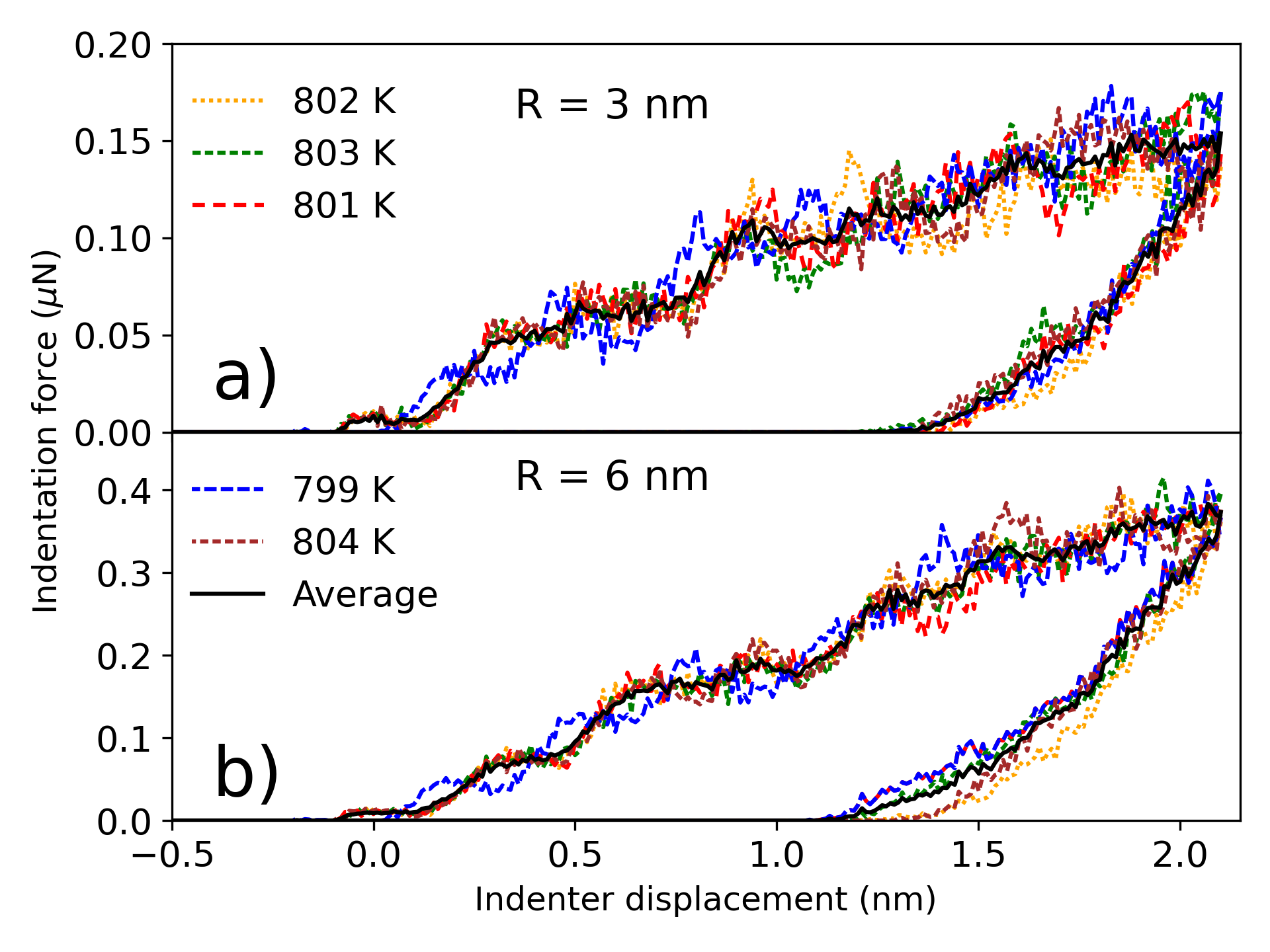}
   \caption{(Color on-line). P-h curve of the nanoindentation of [001] 
   Mo at 800 K 
   by considering different thermalized samples for an indenter tip of 
   3 nm in a) and 
   6 nm in b). The average of the Load as a function of the indention
   depth is 
   included and computed as $\langle $P$(h) \rangle 
   = 1/N \sum^N_k $P$(h)_k$ with N=5 as the number of the MD simulations.
   }
   \label{fig:800KPhcurve}
\end{figure}

In Fig. \ref{fig:ExpCompHardness} we present the temperature dependence 
of hardness of Mo on the [001] orientation at 2.1 nm depth 
and its comparison
to experimental data reported by Plummer et al. \cite{PlummerOxford} 
for the same crystal orientation. 
We also include experimental data of Mo by 
Pisarenko et al. \cite{Pisarenko} and 
Minnert et al. \cite{MINNERT2020108727} for polycrystalline 
Mo samples as reference.
We finally include the experimental measurements by Terentyev et al.
\cite{TERENTYEV2020105222} for W showing a monotonically decrease 
of hardness as a function of the temperature.
The high temperature behavior resembles the one observed in the 
experiments \cite{PlummerOxford,Pisarenko,MINNERT2020108727} 
where the material hardness reaches a saturation at a 
temperature of $\sim$ 700 K. This is encouraging, given that the
experimentally studied depths were much larger, pointing towards 
maintaining the same trend in thermal effects, despite the strong 
size effects. This finding is consistent with successfully modified 
Nix-Gao models that suggest that the emergence of thermal effects 
arises only through the statistically stored dislocation density in 
the plastic zone and in a power law form, which is then linearly 
added to the geometrically necessary dislocation
density~\cite{VOYIADJIS2010307,Song:2019aa}.
The considered error bars in hardness are obtained through  
additional MD simulations and computed as 
$\sigma = \sqrt{\sum_i \left(x_i - \langle H 
\rangle \right)/N}$ is the standard deviation, $\langle H \rangle$ 
the average of the MD results, and $N = 5$,  the total number of MD
simulations performed. 
As an example, Fig. \ref{fig:800KPhcurve} shows the 
P-h curves of the nanoindenation of the sample at 800 K for a 
radius of 3nm in a) and 6nm in b)
where the increase of the hardness material is observed.

\begin{figure}[!b]
   \centering
   \includegraphics[width=0.48\textwidth]{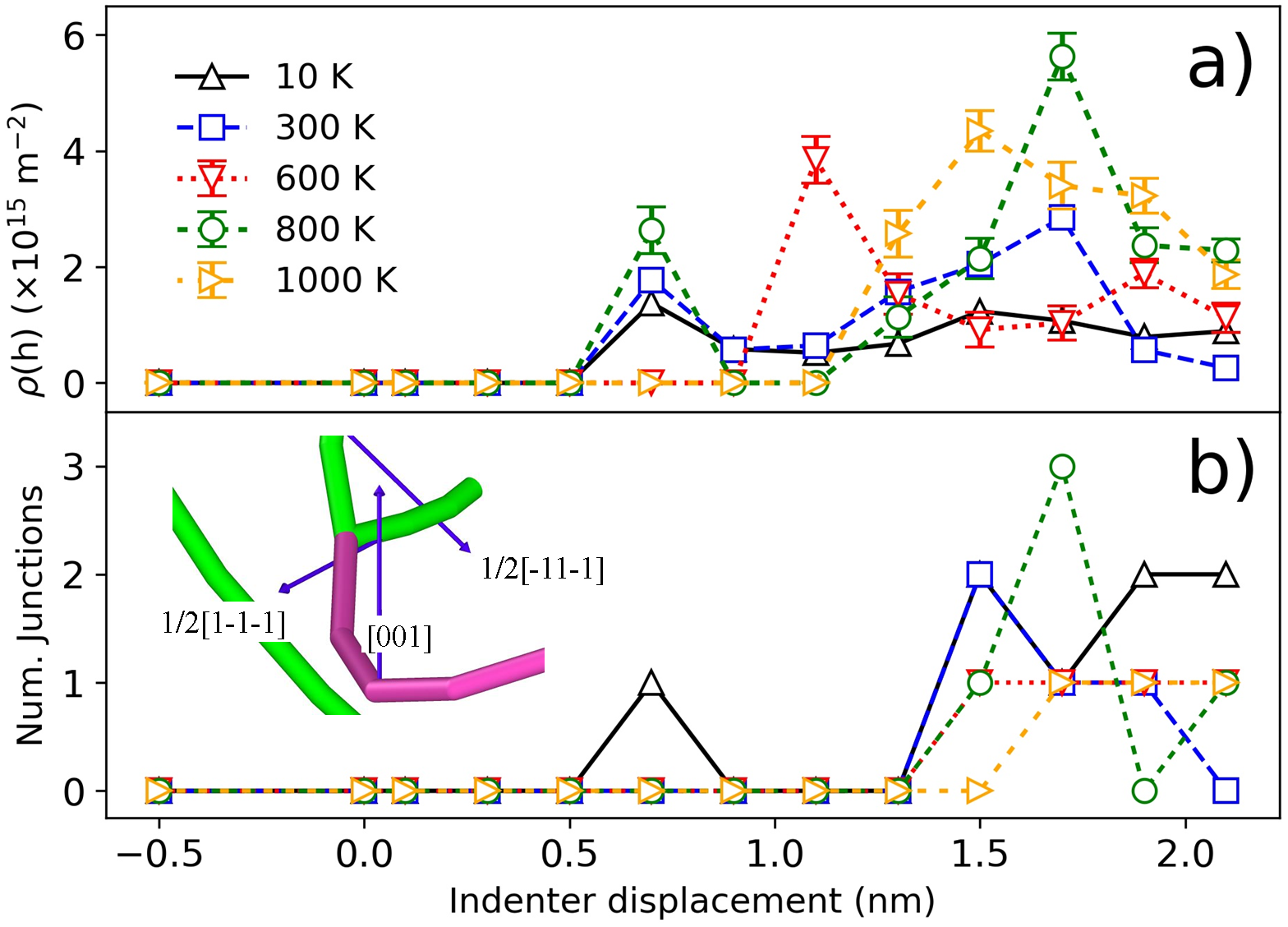}
   \caption{(Color on-line). Dislocation density in a) and 
   number of [001] junctions in b) as function of the indenter displacement 
   for several temperatures, for an indenter radius of 3 nm and a [001]
   crystal orientation.
   Besides the effect of the temperature on the mechanical nanoindentation response, we 
   observed the formation of dislocation junctions in the sample making it to increase its hardness, as shown in b). 
   As an example, we show the dislocation junction in the inset of the 
   figure b) where the [001] dislocation is produced by two 1/2[111] dislocations.
   }
   \label{fig:DislJuncFig}
\end{figure}

The thermomechanical stability seen in the hardness data (see Fig.~\ref{fig:ExpCompHardness}) is strongly correlated with the total dislocation length and dislocation propagation during nanoindentation. 
In Fig. \ref{fig:DislJuncFig}a) we show the dislocation density as function of 
the indenter displacement for different temperatures by considering the 3nm-sized indenter tip. The results for the bigger indenter tip are also presented in the 
supplementary material (\ref{sec:appendix}). 
By using the approximations of a spherical plastic zone, the dislocation density, $\rho($h$)$ is computed as 
\begin{equation}
    \rho(\rm h) = {\rm L}_{\rm T} \left[\frac{2 \pi}{3} \left( R_{pl}^3 - h^3 \right) \right]^{-1},
\end{equation}
where L$_{\rm T}$ is the dislocation length, and $R_{pl}$ is the largest 
distance of a dislocation measured from the indentation 
displacement, considering a hemispherical geometry. 
From our MD simulations, we observe (see Fig.~\ref{fig:DislJuncFig}a) 
that the nucleation of  $b = 1/2 [111]$ and $b=[001]$ dislocations is
responsible for the hardness saturation at high temperatures. 
 Moreover, as shown in the inset of Fig.~\ref{fig:DislJuncFig}b), the 
 [001] dislocation segments are actually, dislocation junctions between
 1/2[1-1-1] and 1/2[-11-1] dislocations, as well as their symmetry
 equivalents. 
 The number of this type of junctions found in our MD simulations as 
 a function of the indenter displacement is presented in Fig.
 \ref{fig:DislJuncFig}b) and its schematic is displayed in the inset. 
 Here Burgers vectors are depicted as blue arrows and dislocation lines
 are  shown by green and pink lines for 1/2[111] and [001], respectively. 
 In the supplementary material, we also show a visualization of  
dislocation nucleation as function of time for the [001] Mo 
sample at different temperatures to demonstrate the generic formation 
of [001] junctions (\ref{sec:appendix}). 
 These junctions are proliferating at high temperatures, and are highly 
 stable at the maximum load. [001] junctions have been long considered 
 as a major hardening mechanism in BCC metals~\cite{Argon:2008aa} that
 may explain typical strengthening~\cite{PhysRevLett.121.085501}. 

Thus, we suggest that [001] junctions may be responsible for the persistent
thermomechanical stability of Mo single crystals. 
Thermally stable [001] junctions are found to be responsible for 
increasing  material nanohardness, and in contrast, this kind of 
junction formation has been shown to be unstable in other BCC metals,
such as W and Ta~\cite{TERENTYEV2020105222,PhysRevLett.109.075502}. 
We conclude that the combination of kinetics and energetics in 
molybdenum leads to an increased stability of [001] junctions that 
are well known to play key role in strengthening effects of BCC
metals~\cite{Argon:2008aa}.

\section{Concluding remarks}
\label{sec:Concl.}
In this work, we performed MD simulations to investigate 
the thermal stability of the mechanical response of crystalline 
molybdenum during nanoindentation, investigating the effects of
crystal orientation, temperature and dislocation mechanisms.  
We characterized the nanoindentation process in molybdenum, in 
connection to experimental findings, and through tracking strain 
accumulation and dislocation mechanisms for several temperatures, 
sample orientations and indenter tips.
Our simulation results suggest that crystal plasticity in nanoindentation 
of molybdenum follows closely known defect nucleation mechanisms for 
BCC metals, albeit with distinct differences in the formation and 
thermal stability of [001] junctions, that are observed and quantified 
as function of  temperature during loading. The observed [001] junction 
formation in molybdenum was shown to be responsible for increasing 
the nano-hardness of  molybdenum, by correlating it to the dislocation
density of each dislocation type. In contrast, such stable junction 
formation  is barely observed during  nanoindentation of W and
Ta~\cite{TERENTYEV2020105222,PhysRevLett.109.075502}. 
Given that [001] junctions are well known culprits of hardening 
in BCC metals~\cite{Argon:2008aa}, we believe that persistent 
high-temperature hardness in Mo may be attributed to [001] 
junction formation. Therefore, the simulation results presented 
in this work suggest that there may be fundamental dislocation-based 
kinetic reasons for molybdenum being  a good, low-maintenance material 
candidate for applications in extreme environmental conditions, 
over other BCC metals.

\section*{Acknowledgments}
We would like to thank {\L}ukasz Kurpaska for inspiring conversations.
 We acknowledge support from the European Union Horizon 2020 research
 and innovation program under grant agreement no. 857470 and from the 
 European Regional Development Fund via the Foundation for Polish 
 Science International Research Agenda PLUS program grant 
 No. MAB PLUS/2018/8. We acknowledge the computational resources 
 provided by the High Performance Cluster at the National Centre 
 for Nuclear Research in Poland,  and also the Seawulf institutional 
 cluster at the Institute for Advanced Computational Science in 
 Stony Brook University.

\appendix
\section{Supplementary material}
\label{sec:appendix}

We provide the analysis of the indented Mo sample for the 
[110] and [111] at different temperatures and indenter size. 
As well as the visualization of the dislocation loops 
at the maximum indentation depth and after unloading process.
\bibliographystyle{iopart-num}
\bibliography{refs,refs2}	

\providecommand{\newblock}{}
\begin{thebibliography}{10}
\expandafter\ifx\csname url\endcsname\relax
  \def\url#1{{\tt #1}}\fi
\expandafter\ifx\csname urlprefix\endcsname\relax\def\urlprefix{URL }\fi
\providecommand{\eprint}[2][]{\url{#2}}

\bibitem{KeMin}
Xue K~M, Wang Z, Wang X, Zhou Y~F and Li P 2020 {\em Materials Science and
  Technology\/} {\bf 0} 1--9

\bibitem{2010M2009375}
Lee S, Edalati K and Horita Z 2010 {\em Materials Transactions\/} {\bf 51}
  1072--1079

\bibitem{HOLLANG2001233}
Hollang L, Brunner D and Seeger A 2001 {\em Materials Science and Engineering:
  A\/} {\bf 319-321} 233--236

\bibitem{Litasov}
Litasov K~D, Dorogokupets P~I, Ohtani E, Fei Y, Shatskiy A, Sharygin I~S,
  Gavryushkin P~N, Rashchenko S~V, Seryotkin Y~V, Higo Y, Funakoshi K,
  Chanyshev A~D and Lobanov S~S 2013 {\em Journal of Applied Physics\/} {\bf
  113} 093507

\bibitem{doi:10.1063/1.324094}
Ragan~III C~E, Silbert M~G and Diven B~C 1977 {\em Journal of Applied
  Physics\/} {\bf 48} 2860--2870

\bibitem{CIESZYKOWSKA2017124}
Cieszykowska I, Janiak T, Barcikowski T {\em et~al.\/} 2017 {\em Applied
  Radiation and Isotopes\/} {\bf 124} 124--131 ISSN 0969-8043

\bibitem{LI2018367}
Li P, Lin Q, Wang X, Tian Y and Xue K~M 2018 {\em International Journal of
  Refractory Metals and Hard Materials\/} {\bf 72} 367--372

\bibitem{doi:10.1098/rspa.1980.0043}
Kumar A, Eyre B~L and Christian J~W 1980 {\em Proceedings of the Royal Society
  of London. A. Mathematical and Physical Sciences\/} {\bf 370} 431--458

\bibitem{BUDD1990129}
Budd M 1990 {\em Journal of Nuclear Materials\/} {\bf 170} 129--133

\bibitem{IterDivertor}
The {D}ivertor \url{https://www.iter.org/mach/divertor} accessed: 15/2/2021

\bibitem{Smith:2003aa}
Smith R, Christopher D, Kenny S~D, Richter A and Wolf B 2003 {\em Physical
  Review B\/} {\bf 67} 245405

\bibitem{Durst:2006aa}
Durst K, Backes B, Franke O and G{\"o}ken M 2006 {\em Acta Materialia\/} {\bf
  54} 2547--2555

\bibitem{Stelmashenko:1993aa}
Stelmashenko N, Walls M, Brown L and Milman Y~V 1993 {\em Acta Metallurgica et
  Materialia\/} {\bf 41} 2855--2865

\bibitem{Syed-Asif:1997aa}
Syed~Asif S and Pethica J 1997 {\em Philosophical Magazine A\/} {\bf 76}
  1105--1118

\bibitem{Bahr:1998aa}
Bahr D~F, Kramer D~E and Gerberich W 1998 {\em Acta materialia\/} {\bf 46} 3605

\bibitem{Kramer:2001aa}
Kramer D, Yoder K and Gerberich W 2001 {\em Philosophical Magazine A\/} {\bf
  81} 2033--2058

\bibitem{Biener:2007aa}
Biener M~M, Biener J, Hodge A~M and Hamza A~V 2007 {\em Physical Review B\/}
  {\bf 76} 165422

\bibitem{PlummerOxford}
Plummer K~P 2021 {\em The Temperature Dependence of Plasticity in Molybdenum\/}
  {PhD} dissertation University of Oxford

\bibitem{VOYIADJIS2010307}
Voyiadjis G~Z, Almasri A~H and Park T 2010 {\em Mechanics Research
  Communications\/} {\bf 37} 307--314

\bibitem{TERENTYEV2020105222}
Terentyev D, Xiao X, Lemeshko S, Hangen U and Zhurkin E 2020 {\em International
  Journal of Refractory Metals and Hard Materials\/} {\bf 89} 105222

\bibitem{BEAKE201863}
Beake B~D and Goel S 2018 {\em International Journal of Refractory Metals and
  Hard Materials\/} {\bf 75} 63--69 ISSN 0263-4368
  \urlprefix\url{https://www.sciencedirect.com/science/article/pii/S026343681830060X}

\bibitem{Pisarenko}
Pisarenko G, Borisenko V and Kashtalyan Y 1964 {\em Powder Metall Met Ceram.\/}
  {\bf 1} 371

\bibitem{BRAUN2019104999}
Braun J, Kaserer L, Stajkovic J, Leitz K~H, Tabernig B, Singer P, Leibenguth P,
  Gspan C, Kestler H and Leichtfried G 2019 {\em International Journal of
  Refractory Metals and Hard Materials\/} {\bf 84} 104999

\bibitem{PhysRevB.66.094110}
Luo W, Roundy D, Cohen M~L and Morris J~W 2002 {\em Phys. Rev. B\/} {\bf 66}(9)
  094110

\bibitem{RPicu}
Picu R 2000 {\em Journal of Computer-Aided Materials Design\/} {\bf 7} 77

\bibitem{10.1007/978-981-10-4109-9_34}
Hu J, Li M, Wang W and Li L 2018 Molecular dynamics simulations on
  nanoindentation experiment of single-layer mos2 circular nanosheets {\em
  Advanced Mechanical Science and Technology for the Industrial Revolution
  4.0\/} ed Yao L, Zhong S, Kikuta H, Juang J~G and Anpo M (Singapore: Springer
  Singapore) pp 333--339 ISBN 978-981-10-4109-9

\bibitem{PhysRevLett.109.075502}
Alcal\'a J, Dalmau R, Franke O, Biener M, Biener J and Hodge A 2012 {\em Phys.
  Rev. Lett.\/} {\bf 109}(7) 075502
  \urlprefix\url{https://link.aps.org/doi/10.1103/PhysRevLett.109.075502}

\bibitem{Remington:2014aa}
Remington T, Ruestes C~J, Bringa E~M, Remington B~A, Lu C, Kad B and Meyers M~A
  2014 {\em Acta Materialia\/} {\bf 78} 378--393

\bibitem{Sato:2019aa}
Sato Y, Shinzato S, Ohmura T and Ogata S 2019 {\em International Journal of
  Plasticity\/} {\bf 121} 280--292
  \urlprefix\url{https://www.sciencedirect.com/science/article/pii/S0749641919302335}

\bibitem{Kaufmann:2013aa}
Kaufmann D, Schneider A~S, M{\"o}nig R, Volkert C~A and Kraft O 2013 {\em
  International Journal of Plasticity\/} {\bf 49} 145--151
  \urlprefix\url{https://www.sciencedirect.com/science/article/pii/S0749641913000740}

\bibitem{Kaufmann:2011aa}
Kaufmann D, M{\"o}nig R, Volkert C~A and Kraft O 2011 {\em International
  Journal of Plasticity\/} {\bf 27} 470--478
  \urlprefix\url{https://www.sciencedirect.com/science/article/pii/S0749641910001130}

\bibitem{Papanikolaou:2017aa}
Papanikolaou S, Cui Y and Ghoniem N 2017 {\em Modelling and Simulation in
  Materials Science and Engineering\/} {\bf 26} 013001

\bibitem{cryst7100321}
Voyiadjis G~Z and Yaghoobi M 2017 {\em Crystals\/} {\bf 7} 321

\bibitem{ZHAO2018365}
Zhao J, Huang P, Xu K, Wang F and Lu T 2018 {\em Thin Solid Films\/} {\bf 653}
  365--370

\bibitem{MINNERT2020108727}
Minnert C, Oliver W~C and Durst K 2020 {\em Materials \& Design\/} {\bf 192}
  108727

\bibitem{WHEELER2015354}
Wheeler J, Armstrong D, Heinz W and Schwaiger R 2015 {\em Current Opinion in
  Solid State and Materials Science\/} {\bf 19} 354--366

\bibitem{10.1088/1361-651X/abf152}
Dom\'inguez-Guti\'errez F~J, Byggm\"astar J, Nordlund K, Djurabekova F and von
  Toussaint U 2021 {\em Modelling and Simulation in Materials Science and
  Engineering\/}
  \urlprefix\url{http://iopscience.iop.org/article/10.1088/1361-651X/abf152}

\bibitem{PLIMPTON19951}
Plimpton S 1995 {\em Journal of Computational Physics\/} {\bf 117} 1 -- 19

\bibitem{2007MJ200769}
Shimizu F, Ogata S and Li J 2007 {\em Materials Transactions\/} {\bf 48}
  2923--2927

\bibitem{Lee}
Lee S, Vaid A, Im J {\em et~al.\/} 2020 {\em Nature Communications\/} {\bf 11}
  2367

\bibitem{Christopher_2001}
Christopher D, Smith R and Richter A 2001 {\em Nanotechnology\/} {\bf 12}
  372--383 \urlprefix\url{https://doi.org/10.1088/0957-4484/12/3/328}

\bibitem{Frenkel:2001aa}
Frenkel D and Smit B 2001 {\em Understanding molecular simulation: from
  algorithms to applications\/} vol~1 (Elsevier) ISBN 0080519989

\bibitem{AcklandGJ}
Ackland G~J and Thetford R 1987 {\em Philosophical Magazine A\/} {\bf 56}
  15--30

\bibitem{Salonen_2003}
Salonen E, J\"arvi T, Nordlund K and Keinonen J 2003 {\em Journal of Physics:
  Condensed Matter\/} {\bf 15} 5845--5855

\bibitem{oliver_pharr_1992}
Oliver W and Pharr G 1992 {\em Journal of Materials Research\/} {\bf 7}
  1564–1583

\bibitem{ovito}
Stukowski A {2010} {\em {Modelling and simulation in materials science and
  engineering}\/} {\bf {18}} ISSN {0965-0393}

\bibitem{GOEL2015249}
Goel S, Beake B, Chan C~W, {Haque Faisal} N and Dunne N 2015 {\em Materials
  Science and Engineering: A\/} {\bf 627} 249--261

\bibitem{Bolin:2019aa}
Bolin R, Yavas H, Song H, Hemker K~J and Papanikolaou S 2019 {\em Crystals\/}
  {\bf 9} 652

\bibitem{Song:2019aa}
Song H, Yavas H, Van~der Giessen E and Papanikolaou S 2019 {\em Journal of the
  Mechanics and Physics of Solids\/} {\bf 123} 332--347

\bibitem{Argon:2008aa}
Argon A 2008 {\em Strengthening mechanisms in crystal plasticity\/} vol~4
  (Oxford University Press on Demand) ISBN 0198516002

\bibitem{PhysRevLett.121.085501}
Sills R~B, Bertin N, Aghaei A and Cai W 2018 {\em Phys. Rev. Lett.\/} {\bf
  121}(8) 085501
  \urlprefix\url{https://link.aps.org/doi/10.1103/PhysRevLett.121.085501}

\end{thebibliography}


\section*{Supplementary material}

# Nanoindentation of single crystalline Mo: Atomistic defect nucleation mechanisms and thermomechanical stability

F. J. Domínguez-Gutiérrez[1,2]*, S. Papanikolaou[1], A. Esfandiarpour[1], P. Sobkowicz[1], M. Alava[1,3]

[1]NOMATEN Centre of Excellence, National Center for Nuclear Research, ul. A. Sołtana 7, 05-400 Otwock, Poland
[2]Institute for Advanced Computational Science, Stony Brook University, Stony Brook, NY 11749, USA
[3]Department of Applied Physics, Aalto University, P.O. Box 11100, FI-00076 Aalto, Espoo, Finland.

We report the analysis of indented Mo samples at the maximum penetration depth for [110] and [111] orientations. We also provide a visualization of the formed dislocations at the maximum penetration depth and after nanoindentation.

In Fig. 1 we show load- indentation depth curves at 10 K for different crystal orientation. The [001] orientation favors the first pop-in load followed by [110] and [111] orientations, for both indenter sizes. The results are in good agreement with reported results by J. Alcala et al. for BCC Ta at 77K.

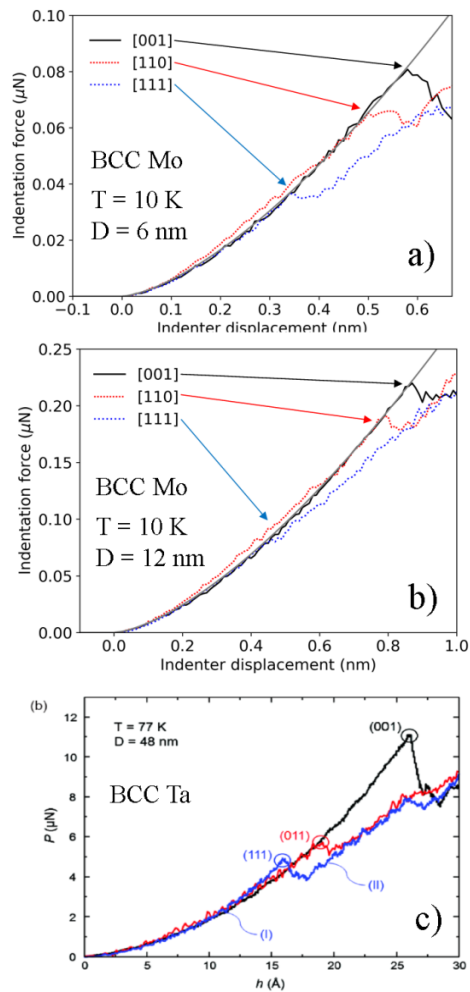

**Figure 1. Load indenter displacement curves for Mo samples at 10 K with different crystal orientations. Comparison to results by J. Alcala et al Phys. Rev. Lett. 109, 075502 (2012) for BCC Ta reaching a good agreement.**

## 1. [111] orientation and indenter ratio of 3 nm.

In Fig. 1, we present the displacement magnitude in a), the atomic shear strain in b) and stress in c) as a function of temperature for [111] orientation. By considering a small indenter tip of 3 nm, we observe that pile ups are more observable by raising temperature. It is also observed that atomic shear strain increases underneath the indenter for high temperature, so do the stress at the maximum indentation depth.

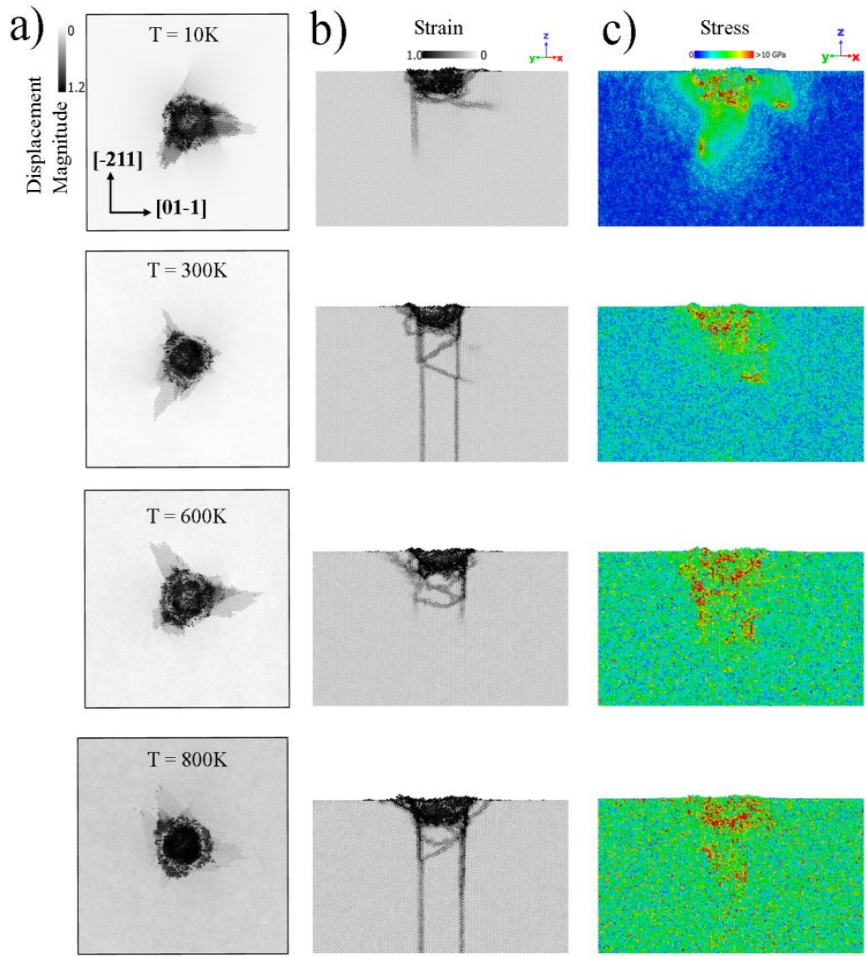

**Figure 2. Crystal orientation [111], Dislocation loops for [111] orientation with a R=3nm indenter.**

In Fig. 2, we show the formed dislocation loops at the maximum indentation depth and after nanoindentation. Noticing that the formation of prismatic loops is more favorable to be formed by increasing the sample temperature. This type of dislocation is preserved after unloading process. The plastic recovery property of Mo is observed at high temperatures where the loops are adsorbed by the surface after nanoindenation showing a decrease in the hardness material. Then, the formation of a prismatic loop suggests the increase of the hardness at 800 K.

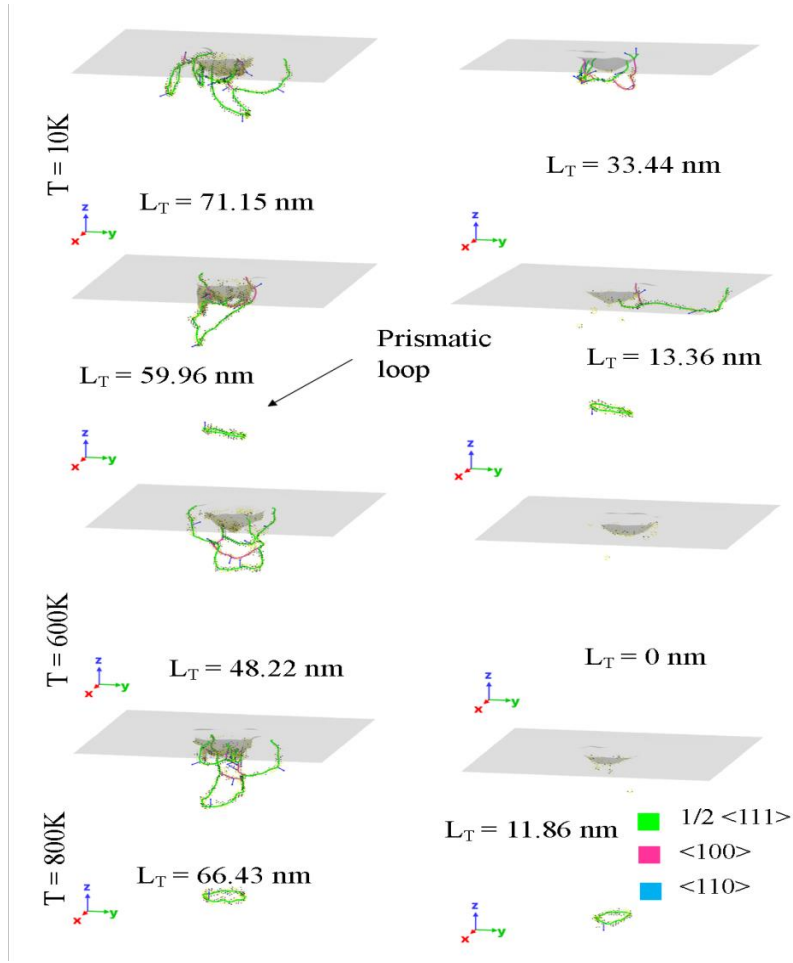

**Figure 3. Formed dislocations at the maximum indentation depth (right side) and after nanoindentation (left side) as a function of the temperature.**

## 2. [110] orientation and indenter ratio of 3 nm.

In Fig. 3, we show the distribution of atomic displacement magnitude at different temperatures, as well as the atomic shear strain and von Mises stress. The effect of temperature on the mechanical response of Mo at this crystal orientation is observed in the results for atomic strain where propagation of the dislocations decreases as temperature rises.

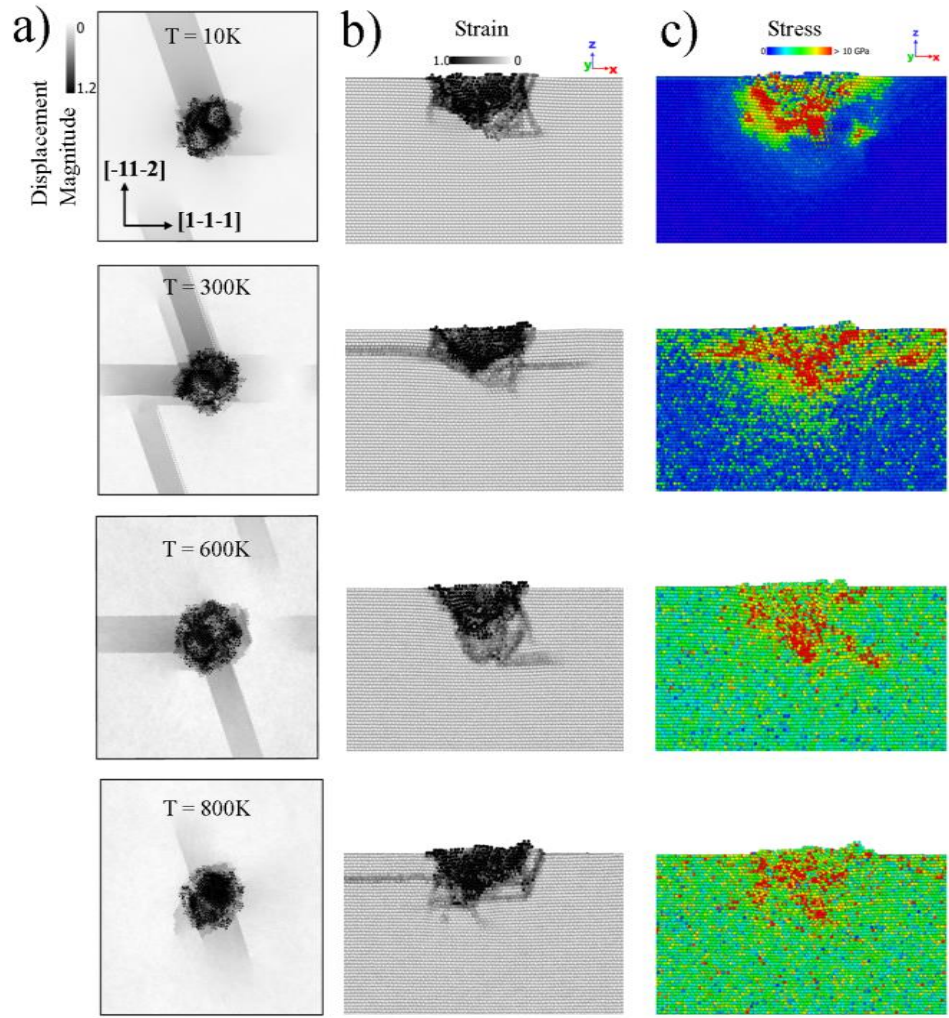

**Figure 4. Results for Mo on the crystal orientation [110] with a R = 3 nm indenter. Displacement magnitude, atomic shear strain and von Mises stress is reported at the maximum indentation depth.**

In Fig. 4, we show a visualization of formed dislocations at the maximum indentation depth and also, after nanoindentation unloading. The effect of temperature is mainly the increase of the length of dislocations. For this crystal orientation the hardness decreases monotonically and formation of prismatic loops is not observed.

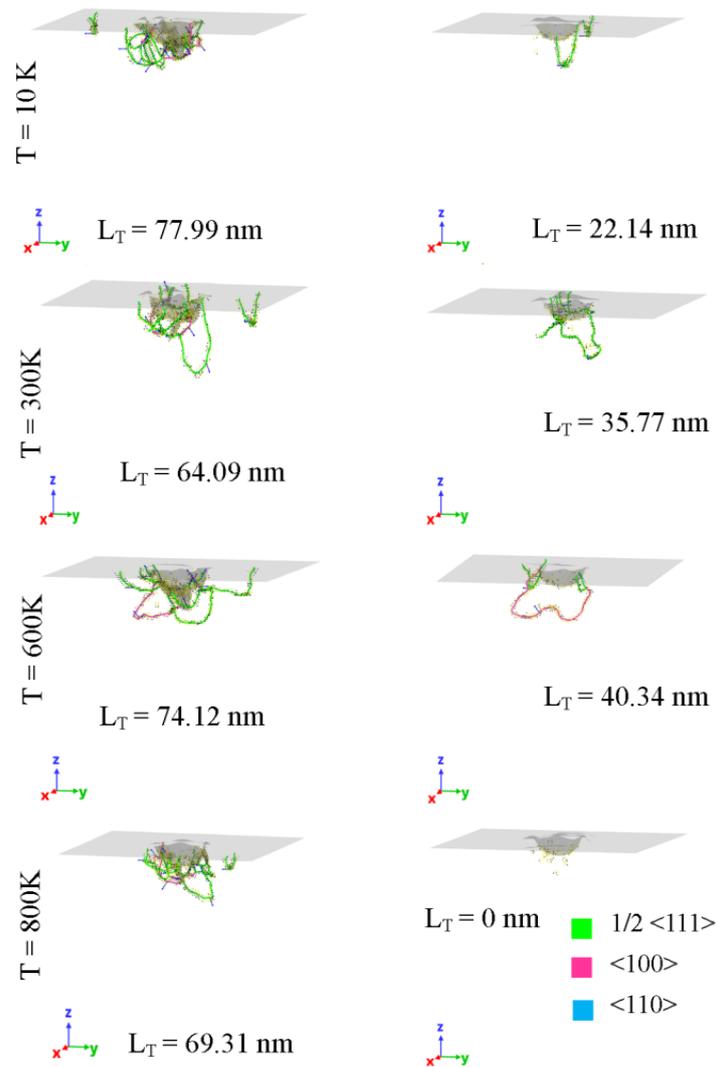

**Figure 5.** Formed dislocations at the maximum indentation depth (right side) and after nanoindentation (left side) as function of temperature, for the [110] orientation.

## 3. [111] orientation and indenter ratio of 6 nm.

In Fig. 5 we present results for the mechanical nanoindentation response of molybdenum samples oriented [111] by using an indenter tip of radius 6nm. The distribution of displacement magnitude, atomic strain and stress are presented at the maximum indentation depth. The size of the indenter tends to create prismatic loops at every sample temperature. Besides, the pile ups observed at 10 K tend to be more diffused at high temperatures. The atomic strain shows the formation of more dislocations as function of temperature. So does the stress figures, where the formation of prismatic loops is observed. However, thermal mechanisms are intensely activated at high temperature, thus making the visualization of the prismatic loops difficult.

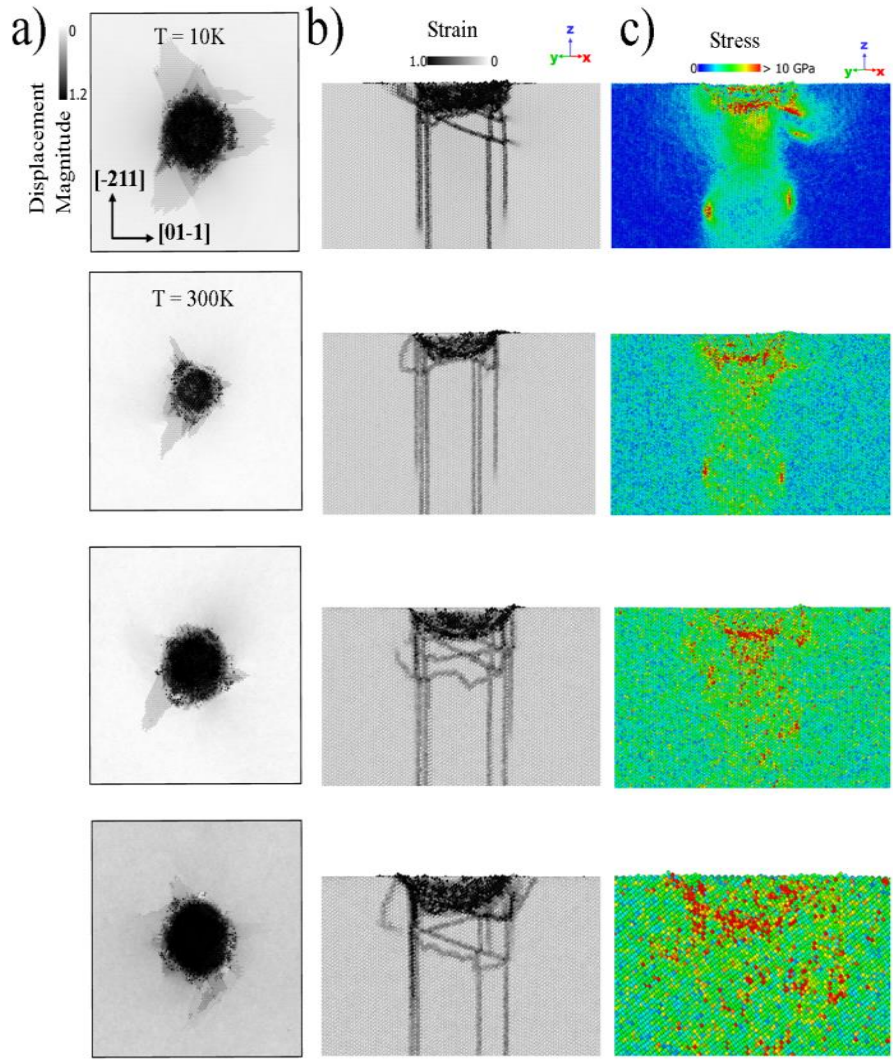

**Figure 6** Results for Mo on the crystal orientation [110] with a R = 6 nm indenter. Displacement magnitude, atomic shear strain and von Mises stress is reported at the maximum indentation depth.

In Fig. 6, we show dislocations formed at the maximum indentation depth for the loading process on the left side panel of the figure and for the unloading process at the right side panel of the figure. We observed that the first dislocation loop formed during the loading process remains after nanoindentation, regardless of the sample temperature. It is worth noticing that the FCC atoms (depicted as yellow spheres) are found around the prismatic loops which can be attributed to the decrease of the hardness of the sample.

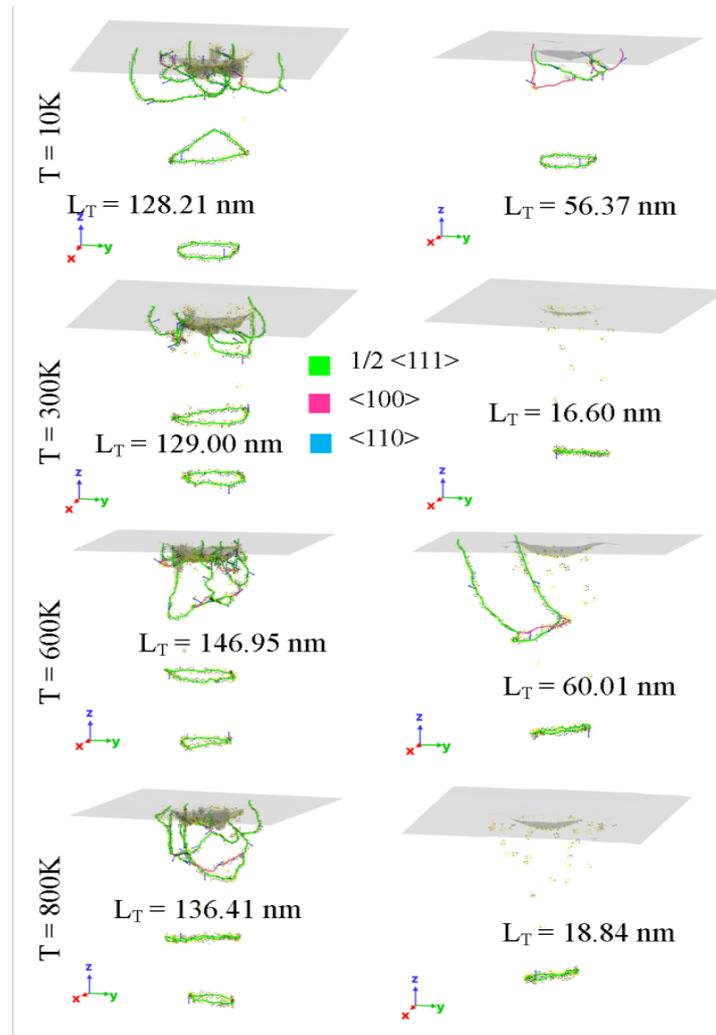

**Figure 7** Formed dislocations at the maximum indentation depth (right side) and after nanoindentation (left side) as a function of the temperature, for the [111] orientation

### 4. [110] orientation and indenter ratio of 6 nm.

The analysis of the distribution of displacement magnitude, atomic strain, and stress is shown in Fig. 7. The effect of temperature is observed by these results where the atomic strain is higher for the region underneath the indenter tip. The stress of the sample at the maximum indentation depth and different temperatures shows that the sample can nucleate more dislocations.

In Fig. 8, we present the dislocations formed by using DXA, the formation of prismatic loops are observed due to the big size of the indenter during the loading process. After nanoindentation, the prismatic loops remain, as observed for the rest of the crystal orientation and regardless of the sample temperature. The CAT analysis show the FCC atoms around the loops are observed for the other crystal orientations.

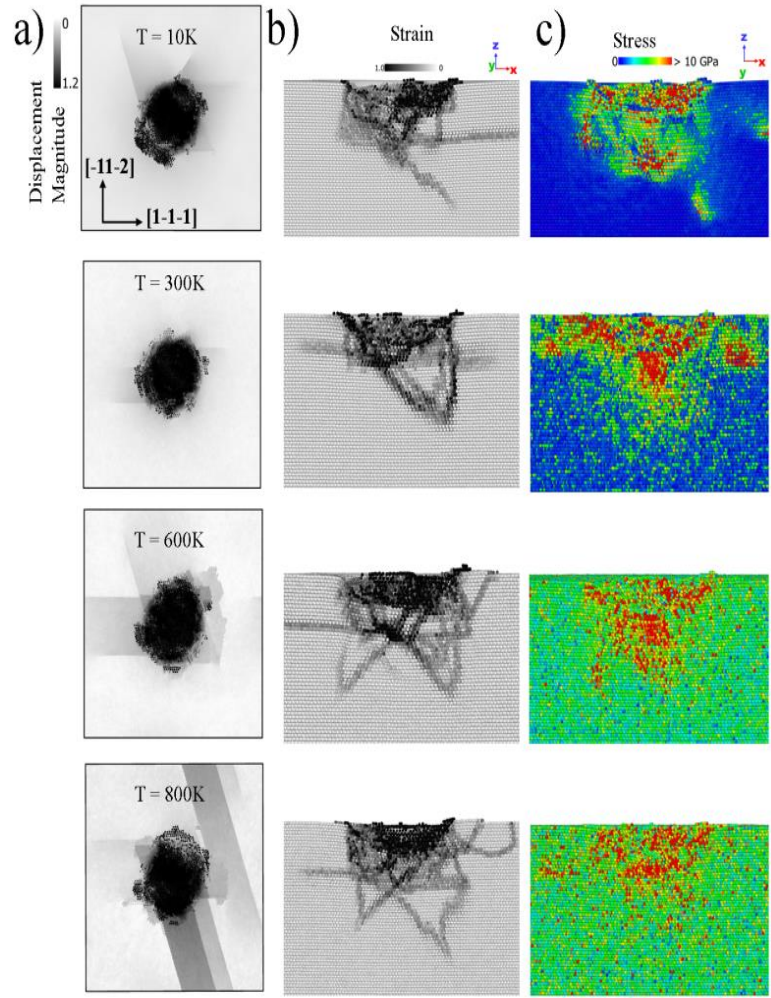

**Figure 8 Results for Mo on the crystal orientation [110] with a R = 6 nm indenter. Displacement magnitude, atomic shear strain and von Mises stress is reported at the maximum indentation depth**

In Fig. 9, we present the dislocations formed at the maximum indentation depth for the loading process on the left side panel of the figure and for the unloading process at the right side panel of the figure. Prismatic loops are formed at all sample temperatures and remain after nanoindentation.

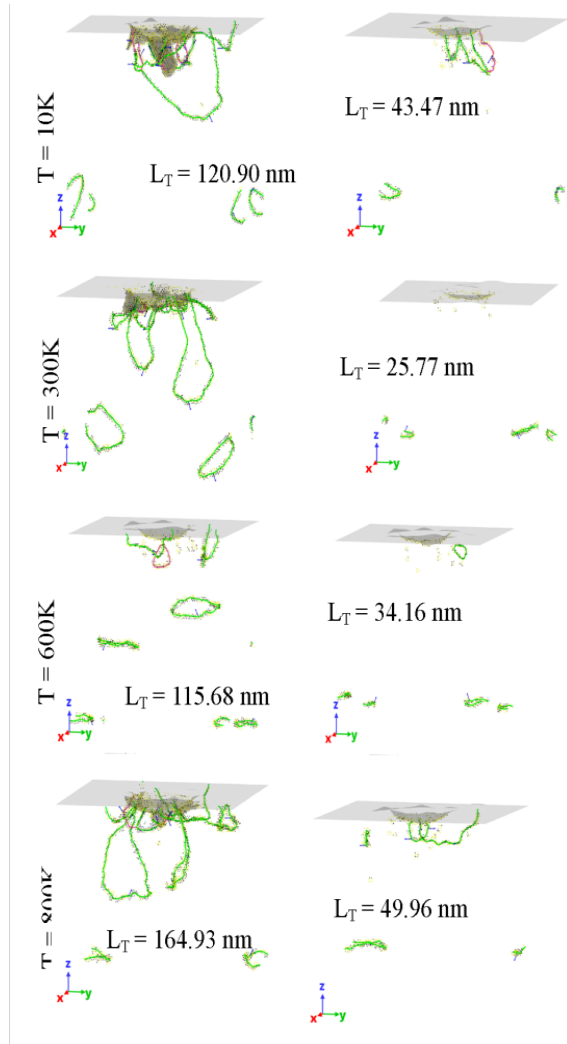

**Figure 9. Formed dislocations at the maximum indentation depth (right side) and after nanoindentation (left side) as a function of the temperature, for the [110] orientation**

[30] Kaufmann D, Schneider A S, Mönig R, Volkert C A and Kraft O 2013 *International Journal of Plasticity* **49** 145–151 URL https://www.sciencedirect.com/science/article/pii/S0749641913000740
[31] Kaufmann D, Mönig R, Volkert C A and Kraft O 2011 *International Journal of Plasticity* **27** 470–478 URL https://www.sciencedirect.com/science/article/pii/S0749641910001130
[32] Papanikolaou S, Cui Y and Ghoniem N 2017 *Modelling and Simulation in Materials Science and Engineering* **26** 013001
[33] Voyiadjis G Z and Yaghoobi M 2017 *Crystals* **7** 321
[34] Zhao J, Huang P, Xu K, Wang F and Lu T 2018 *Thin Solid Films* **653** 365–370
[35] Minnert C, Oliver W C and Durst K 2020 *Materials & Design* **192** 108727
[36] Wheeler J, Armstrong D, Heinz W and Schwaiger R 2015 *Current Opinion in Solid State and Materials Science* **19** 354–366
[37] Domínguez-Gutiérrez F J, Byggmästar J, Nordlund K, Djurabekova F and von Toussaint U 2021 *Modelling and Simulation in Materials Science and Engineering* URL http://iopscience.iop.org/article/10.1088/1361-651X/abf152
[38] Plimpton S 1995 *Journal of Computational Physics* **117** 1 – 19
[39] Shimizu F, Ogata S and Li J 2007 *Materials Transactions* **48** 2923–2927
[40] Lee S, Vaid A, Im J *et al.* 2020 *Nature Communications* **11** 2367
[41] Christopher D, Smith R and Richter A 2001 *Nanotechnology* **12** 372–383 URL https://doi.org/10.1088/0957-4484/12/3/328
[42] Frenkel D and Smit B 2001 *Understanding molecular simulation: from algorithms to applications* vol 1 (Elsevier) ISBN 0080519989
[43] Ackland G J and Thetford R 1987 *Philosophical Magazine A* **56** 15–30
[44] Salonen E, Järvi T, Nordlund K and Keinonen J 2003 *Journal of Physics: Condensed Matter* **15** 5845–5855
[45] Oliver W and Pharr G 1992 *Journal of Materials Research* **7** 1564–1583
[46] Stukowski A 2010 *Modelling and simulation in materials science and engineering* **18** ISSN 0965-0393
[47] Goel S, Beake B, Chan C W, Haque Faisal N and Dunne N 2015 *Materials Science and Engineering: A* **627** 249–261
[48] Bolin R, Yavas H, Song H, Hemker K J and Papanikolaou S 2019 *Crystals* **9** 652
[49] Song H, Yavas H, Van der Giessen E and Papanikolaou S 2019 *Journal of the Mechanics and Physics of Solids* **123** 332–347
[50] Argon A 2008 *Strengthening mechanisms in crystal plasticity* vol 4 (Oxford University Press on Demand) ISBN 0198516002
[51] Sills R B, Bertin N, Aghaei A and Cai W 2018 *Phys. Rev. Lett.* **121**(8) 085501 URL https://link.aps.org/doi/10.1103/PhysRevLett.121.085501


\end{document}